\documentclass[pra,twocolumn,showpacs,amsmath,amssymb,superscriptaddress]{revtex4-1}

\usepackage{color}
\usepackage{graphicx}
\usepackage{float}
\usepackage{bm}
\usepackage{commath}
\usepackage{color}
\usepackage{amsmath}

\def\be{\begin{equation}}
\def\ee{\end{equation}}

\begin{document}

\title{Operation and intrinsic error budget of a two-qubit cross-resonance gate}

\author{Vinay Tripathi}

\affiliation{Department of Electrical and Computer Engineering, University of California, Riverside, CA 92521, USA}

\author{Mostafa Khezri}

\affiliation{Department of Electrical and Computer Engineering, University of California, Riverside, CA 92521, USA}
\affiliation{Department of Electrical Engineering, University of Southern California, Los Angeles, CA 90089, USA}
\author{Alexander N. Korotkov}

\altaffiliation[Present address: ]{Google Inc., Venice, CA 90291, USA}
\affiliation{Department of Electrical and Computer Engineering, University of California, Riverside, CA 92521, USA}

\date{\today}
\begin{abstract}
We analyze analytically, semi-analytically, and numerically the operation of Cross-Resonance (CR) gate for superconducting qubits (transmons). We find that a relatively simple semi-analytical method gives accurate results for the CNOT-equivalent gate duration and compensating single-qubit rotations. It also allows us to minimize the CNOT gate duration over the amplitude of the applied microwave drive and find dependence on the detuning between the qubits. However, full numerical simulations are needed to calculate the intrinsic fidelity of the CR gate. We decompose numerical infidelity into contributions from various physical mechanisms, thus finding the intrinsic error budget. In particular, at small drive amplitudes, the CR gate fidelity is limited by imperfections of the target-qubit rotations, while at large amplitudes it is limited by leakage. The gate duration and fidelity are analyzed numerically as functions of the detuning between qubits, their coupling, drive frequency, relative duration of pulse ramps, and microwave crosstalk. The effect of the echo sequence is also analyzed numerically. Our results show that the CR gate can provide intrinsic infidelity of less than $10^{-3}$ when a simple pulse shape is used.
\end{abstract}

\maketitle

\section{Introduction}

Two decades have passed since the first superconducting qubit was created \cite{Nakamura1999}, and today superconducting quantum computing is a well-developed field with various types and uses of qubits \cite{Neill2018, Kandala2017, Chou2018, Hong2019, Hacohen-Gourgy2016, Satzinger2018, King2018, Kurpiers2018, Minev2018, Masuyama2018}. Currently, the most popular type of superconducting qubits is the transmon \cite{Koch2007} (including its Xmon  and gmon modifications \cite{Barends2013, Chen2014}), though other types of  qubits (e.g., \cite{Chiorescu2003, Yan2016, Lin2018}) are also of interest. Besides sufficiently good coherence of the qubits, quantum computing applications need high-fidelity gates forming a universal set \cite{Nielsen-Chuang-book}. While single-qubit gates are already considered to be simple and accurate, current fidelity of two-qubit gates also exceeds 99\% \cite{Barends2014, Sheldon2016, Hong2019}.

One of the high-fidelity two-qubit gates used for superconducting qubits is the Cross-Resonance (CR) gate \cite{Rigetti2010, Chow2011}. In this gate,  two frequency-detuned qubits have a fixed coupling (usually via a resonator) and one of them (called control qubit) is driven by a microwave with frequency of the other one (target qubit). This induces Rabi oscillations of the target qubit, whose frequency depends on the state ($|0\rangle$ or $|1\rangle$) of the control qubit, thus entangling the two qubits and providing a natural way to realize CNOT operation. Since the CR gate uses only microwave control, it permits to use single-junction transmons, thus avoiding sensitivity to flux noise. However, the drawback is a relatively long gate duration compared with the gates based on tune-detune operation \cite{Barends2014}.

The idea of the CR gate was proposed in 2006 in Ref.\ \cite{Paraoanu2006} and then experimentally implemented for flux qubits in 2010 in Ref.\ \cite{deGroot2010} under the name of Selective Darkening (the difference compared with a simple CR gate is an additional active cancellation pulse applied to the target qubit). The CR terminology was introduced in 2010 in the theoretical paper \cite{Rigetti2010} and the first experiment under this name was realized with capacitively-shunted flux qubits in 2011 in Ref.\ \cite{Chow2011} with the fidelity of 81\%. In 2012 the CR gate was applied to transmons \cite{Chow2012}, with resulting fidelity of 95\%. Since that time the CR gate was used in numerous experiments by several groups (e.g., \cite{Corcoles2013, Corcoles2015, Takita2016, Sheldon2016, Kandala2017, Ware2015, Naik2019}), with gradual increase of maximum fidelity. An important improvement of the CR operation was achieved by using the echo sequence \cite{Corcoles2013, Takita2016}, which not only increased the fidelity but also allowed protocols that avoid   compensating one-qubit rotations in implementing the CNOT gate. The CR gate with duration of 160 ns and fidelity of 99.1\%   reported in Ref.\ \cite{Sheldon2016} was achieved by using both the echo sequence and active cancellation pulses applied to the target qubit.

In spite of extensive experimental use of the CR gate, its theoretical analysis has been rather limited. Besides the initial papers \cite{Paraoanu2006,  Rigetti2010} outlying the main idea, the CR gate was analyzed in Ref.\ \cite{deGroot2012} with an account of the next level, briefly mentioned in Ref.\ \cite{Gambetta2013conf}, and analyzed in detail in the recent paper \cite{Magesan2018}. There were also numerical studies \cite{Willsch2017, Kirchhoff2018} and related papers \cite{Economou2015,Allen2017}.

In this paper, we analyze the operation of the basic CR gate for transmons  (using a simple pulse shape without the echo sequence, which is considered only in the Appendix) at three levels of complexity and accuracy: analytical, semi-analytical, and numerical. Some of the goals of our analysis are similar to those of Ref.\ \cite{Magesan2018}; however, the approach is very different. After discussing the ideal theory of CR operation for transmons (using the Duffing oscillator model), we develop the next-order approximation somewhat similar to that of Ref.\ \cite{Magesan2018} (which still does not work well, as follows from comparison with full numerics), and then develop the semi-analytical approach, based on the numerical solution of a simple one-qubit time-independent Schr\"odinger equation.

The semi-analytical approach gives very accurate results (compared with full numerics) for the CNOT-equivalent gate duration and compensating single-qubit rotations. In particular, it can be used to find the shortest CNOT gate duration, corresponding optimal drive amplitude, and their dependence on detuning between the qubits. However, the semi-analytical approach cannot be used for finding intrinsic fidelity of the gate (neglecting decoherence), for which we use full numerical simulation. Our numerical simulation includes $7\times 5$ levels in the qubits (we replace qubit coupling via resonator with an equivalent direct coupling) and is based on Magnus expansion \cite{Magnus1954} for the evolution matrices. We use a simple pulse shape with cosine-shaped ramps and a flat middle part.

After calculating the gate infidelity, we numerically decompose it into the contributions from various physical mechanisms, thus finding intrinsic error budget. We show that at small drive amplitudes, the error is dominated by imperfections of the unitary operation within the computational subspace. In contrast, at large drive amplitudes (which correspond to reasonably short gate durations) the infidelity is dominated by leakage. Particular leakage channels depend on detuning between the qubits. In this regime, the analytical estimate for the leakage probability agrees reasonably well with the results for the gate infidelity.

Using numerical results for the CNOT gate duration and fidelity, we analyze their dependence on various parameters, including detuning between qubits, their coupling, drive frequency, smoothness (relative duration of the pulse ramps), and microwave crosstalk. In the Appendix, we also analyze the effect of the echo sequence. Our results show that the CR gate can provide intrinsic infidelity of about $10^{-3}$ (and even less, comparable to $10^{-4}$) with a simple pulse shape.

The paper is organized as follows. In Sec.\ \ref{Sec:System&Ham} we discuss the system and its Hamiltonian. In Sec.\ \ref{sec-analytics} we first consider the ideal operation of the CR gate, then derive the next-order analytics, and then develop the semi-analytical approach. The numerical method is discussed in Sec.\ \ref{sec-numerical}. Numerical results for the CNOT-equivalent gate duration and compensating single-qubit rotations are discussed in Sec.\ \ref{sec-num-CNOT}. Then in Sec.\ \ref{sec-err} we analyze the error budget for the CNOT-gate intrinsic infidelity. In Sec.\ \ref{sec-parameters}, we discuss the dependence of CNOT duration and infidelity on parameters. Finally, we conclude in Sec.\ \ref{sec-conclusion}. In the Appendix, we analyze the echo-CR gate operation.

\label{model}

\section{System and Hamiltonian}
\label{Sec:System&Ham}

\begin{figure}[t]
\hspace{0.2cm} \includegraphics[width=0.75\linewidth]{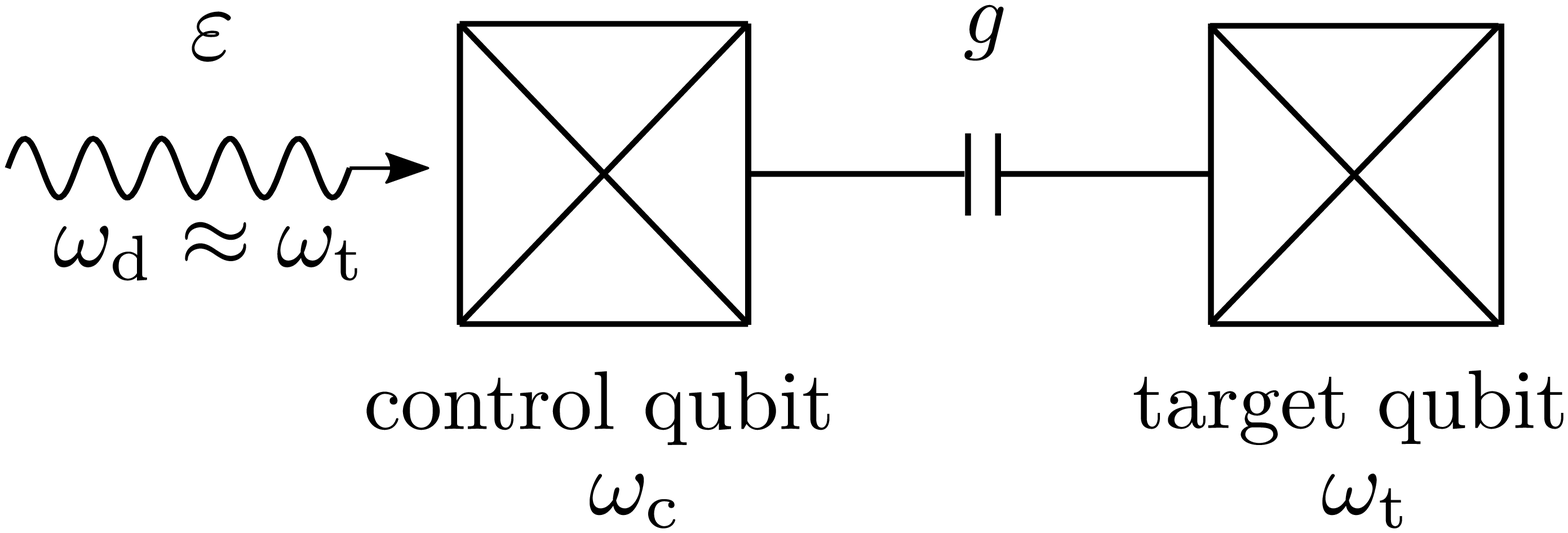}
	\caption{Schematic of the CR gate: detuned control and target qubits (transmons with frequencies $\omega_{\rm c}$ and $\omega_{\rm t}$) have coupling $g$, and the control qubit is microwave-driven at the frequency of the target qubit, $\omega_{\rm d}\approx \omega_{\rm t}$. The microwave drive amplitude is $\varepsilon$.
}
	\label{fig1}
\end{figure}

In the CR gate, the control and target qubits (with frequencies $\omega_{\rm c}$ and $\omega_{\rm t}$, respectively) are usually detuned by 50--300 MHz and are permanently coupled via a resonator. However, in this paper, for simplicity we will consider a direct qubit-qubit coupling $g$ (Fig.\ \ref{fig1}) since the usual analysis of the CR gate \cite{Gambetta2013conf,Magesan2018} also reduces the coupling via a resonator to an effective direct coupling. For the CR operation, the control qubit is rf-driven at the frequency of the target qubit, $\omega_{\rm d}\approx \omega_{\rm t}$. This produces an effective drive ($x$-rotation) of the target qubit, with the strength depending on the state of the control qubit. Such a process can be naturally used to realize the CNOT gate by calibrating the target-qubit rotation angle difference (between rotations for the control-qubit states $|0\rangle$ and $|1\rangle$) to be equal to $\pi$ and somehow compensating the target-qubit rotation for the control-qubit state $|0\rangle$. This compensation can be done, for example, by using the echo sequence \cite{Corcoles2013, Takita2016, Sheldon2016} or active cancellation  \cite{deGroot2010, Sheldon2016}; however, in this paper we will assume that the compensation is done afterward \cite{Rigetti2010, Chow2011} by applying single-qubit rotations (the echo sequence is considered only in the Appendix). We intentionally consider the simplest case in order to focus on developing a good understanding of the basic operation of the CR gate.

\begin{figure}[b]
\hspace{-0.5cm}  \includegraphics[width=0.8\linewidth]{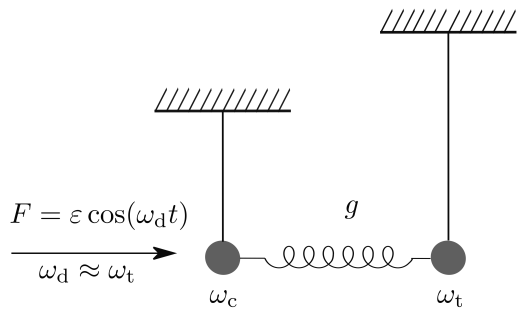}
	\caption{Classical CR gate counterpart: two coupled nonlinear oscillators, with one oscillator driven by a periodic force $F$ on-resonance with the other oscillator.
}
	\label{fig2}
\end{figure}

The operating principle of the CR gate can be understood classically, by replacing qubits with classical oscillators (Fig.\ \ref{fig2}).
Since the drive is off-resonance with the control oscillator  ($\omega_{\rm d} \not\approx \omega_{\rm c}$), it will produce very small forced oscillations at the drive frequency $\omega_{\rm d}$. However, since the target oscillator is on-resonance with this frequency ($\omega_{\rm d} \approx \omega_{\rm t}$), it will still get excited via the coupling $g$ with the control oscillator. Note that if the control oscillator is linear, then its own state (its oscillation with frequency $\omega_{\rm c}$) does not matter because of linearity. However, if the control oscillator is nonlinear, then its effective frequency depends on its own state (i.e., amplitude of $\omega_{\rm c}$-oscillations); therefore, the amplitude of the small forced oscillations of the control oscillator and consequently the excitation rate of the target oscillator will depend on the control-oscillator state. This simple classical picture explains the basic physical mechanism of the CR gate operation for transmons, which are slightly nonlinear oscillators. It also explains why the CR gate speed depends on
nonlinearity of the control qubit and practically does not depend on the target-qubit nonlinearity.

For quantum analysis of the CR gate (Fig.\ \ref{fig1}), let us start with the rotating-frame Hamiltonian (the rotating frame is based on the drive frequency $\omega_{\rm d}$)
    \be
H = H_{\rm qb} + H_g +H_\varepsilon ,
    \label{Ham}\ee
where $H_{\rm qb}$ describes two uncoupled transmon qubits, $H_g$ describes their coupling, and $H_\varepsilon$ describes the microwave drive on the control qubit.

The uncoupled-qubit part can be written as
    \begin{eqnarray}
&& H_{\rm qb}= \sum\nolimits_{n,m}(E_{n}^{\rm (c)}+E_{m}^{\rm (t)}) \, |n,m\rangle \langle n,m|,
    \label{Ham-qb}\\
&& E_{n}^{\rm (c)} = E_{n}^{\rm (c,\,lf)}-n\omega_{\rm d}, \,\,\,
E_{m}^{\rm (t)} = E_{m}^{\rm (t,\,lf)}-m\omega_{\rm d},
    \end{eqnarray}
where in the notation $|n,m\rangle$ the control-qubit state is at the left ($n=0,1,2, ...$) and the target-qubit state is at the right ($m=0,1,2, ...$), the control-qubit energies $E_{n}^{\rm (c)}$ in the rotating frame are related to the laboratory-frame energies $E_{n}^{\rm (c,\, lf)}$ via the drive frequency $\omega_{\rm d}$, and there is a similar relation for the target-qubit energies $E_{m}^{\rm (t)}$. We set $E_{0}^{\rm (c)}=E_{0}^{\rm (t)}=0$.
For the energies $E_{n}^{\rm (c)}$ and $E_{m}^{\rm (t)}$, in this paper we use the Duffing (Kerr) oscillator approximation,
    \begin{eqnarray}
&& E^{\rm (c)}_n = n(\Delta+\delta) - \frac{n(n-1)}{2} \, \eta_{\rm c}, \,\,\,
    \label{E-c-n} \\
&& E_{m}^{\rm (t)} = m\delta - \frac{m(m-1)}{2} \, \eta_{\rm t},
    \label{E-t-m} \\
&& \Delta \equiv \omega_{\rm c} -\omega_{\rm t}, \,\,\,  \delta \equiv \omega_{\rm t}-\omega_{\rm d} \approx 0, \quad
    \label{Delta,delta}\end{eqnarray}
where $\Delta$ is the detuning between the qubits, while $\eta_{\rm c}$ and $\eta_{\rm t}$ are anharmonicities of the control and target qubits, respectively (for transmons $\eta_{\rm c}>0$ and  $\eta_{\rm t}>0$). A small mismatch $\delta$ between the drive frequency $\omega_{\rm d}$ and the {\it bare} frequency $\omega_{\rm t}$ of the target qubit  can be used, e.g., to make the drive exactly resonant with the hybridized target qubit for the control-qubit states $|0\rangle$ or $|1\rangle$ (or in between). Note that $\Delta+\delta = \omega_{\rm c}-\omega_{\rm d}$. Instead of the approximation (\ref{E-c-n})--(\ref{Delta,delta}), it is possible to use numerical results for the transmon energies or at least the improved approximation \cite{Khezri2018, Sank2016}. However, we prefer the simple approximation for easier comparison with the previous theoretical analyses of the CR gate.

The qubit-qubit coupling Hamiltonian $H_g$ in general couples all pairs of the bare states $|n,m\rangle$ and $|n',m'\rangle$. However, in this paper, we use the simplest (traditional) approximation for transmons by keeping only the excitation-preserving terms, i.e., applying the Rotating Wave Approximation (RWA), and using the matrix elements for linear oscillators,
    \be
H_g = \sum\nolimits_{n,m}   g \, \sqrt{nm} \, |n,m-1\rangle \langle n-1, m| + {\rm h.c.},
    \label{Ham-g}\ee
additionally assuming (without loss of generality) that the coupling constant $g$ is real. Similarly, we use the RWA linear-oscillator matrix elements for the drive Hamiltonian (in the rotating frame),
    \be
H_\varepsilon = \sum\nolimits_{n,m} \varepsilon (t) \, \sqrt{n}\, |n,m\rangle \langle n-1,m| + {\rm h.c.},
    \label{Ham-varepsilon}\ee
where the complex amplitude $\varepsilon$ of the drive depends on time, so that $\varepsilon (t)$ is the pulse shape of the CR gate, with $\varepsilon(t)=0$ before and after the gate. Instead of Hamiltonians (\ref{Ham-g}) and (\ref{Ham-varepsilon}), it is possible to use improved perturbative Hamiltonians \cite{Khezri2018, Sank2016} or numerical matrix elements for transmons, but in this paper, we use the simple traditional approximation. Here we do not consider the microwave crosstalk \cite{Chow2011, Chow2012, Sheldon2016, Magesan2018}; however, it will be added in Sec.\ \ref{sec-parameters}.

\begin{figure}[t]
\includegraphics[width=\linewidth]{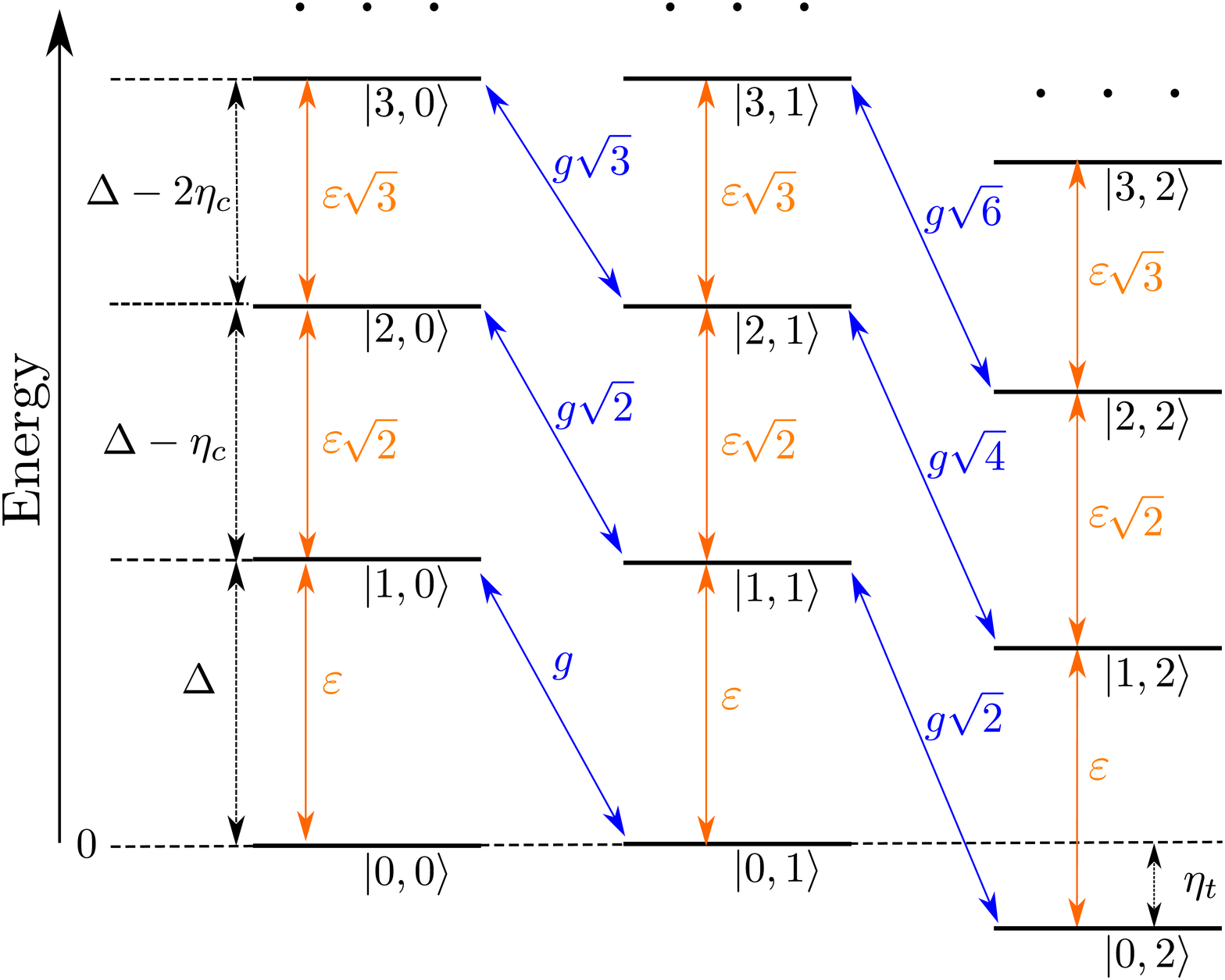}
	\caption{Diagram of bare energy levels for the CR gate: each vertical ladder is for all control-qubit states, with a fixed state of the target qubit (we use notation $|{\rm control,\, target}\rangle$). Slanted blue lines illustrate coupling between the bare levels due to the qubit-qubit coupling $H_g$, orange lines are due to the drive Hamiltonian $H_\varepsilon$. On this diagram we assumed $\delta=0$ (a non-zero $\delta$ would produce an energy shift between the ladders; also, in this case, $\Delta$ should be replaced with $\Delta +\delta$). Control-qubit states above $|3\rangle$ and target-qubit states above $|2\rangle$ are not shown.}
	\label{fig3}
\end{figure}

It is convenient to draw a diagram (Fig.\ \ref{fig3}) of bare levels $|n,m\rangle$, in which the left ladder of levels corresponds to the target-qubit state $|0\rangle$ ($m=0$), the next ladder corresponds to the target-qubit state $|1\rangle$, then $|2\rangle$, and so on. Note that for $\delta =0$, the left two ladders are at exactly equal energies. In Fig.\ \ref{fig3} the coupling $H_g$ is represented by slanted blue arrows and the drive $H_\varepsilon$ corresponds to vertical orange arrows. For clarity, in Fig.\ \ref{fig3} we show the case $\Delta > 2\eta_{\rm c}$, while in experiments usually $0<\Delta < \eta_{\rm c}$. In such a case, all ladders turn down after the states $|1,m\rangle$ and the diagram becomes visually complicated, so for gaining intuition it is easier to use the case of Fig.\ \ref{fig3}.

Besides the bare states $|n,m\rangle$, we will also use the eigenstates of the Hamiltonian $H_{\rm qb}+H_g$ (without the drive), which we denote by an overline: $\overline{|n,m\rangle}$. The coupling $H_g$ affects the qubit frequencies, so instead of the bare frequency $\omega_{\rm t}$ of the target qubit, we have two eigenfrequencies: $\omega_{\rm t}^{\rm c0}$ and $\omega_{\rm t}^{\rm c1}$, depending on the control-qubit state ($|0\rangle$ and $|1\rangle$, respectively). They can be calculated as
    \be
\omega_{\rm t}^{\rm c0}= E_{\overline{|0,1\rangle}}^{\rm (lf)} - E_{\overline{|0,0\rangle}}^{\rm (lf)}\, , \,\,\,\,
\omega_{\rm t}^{\rm c1}= E_{\overline{|1,1\rangle}}^{\rm (lf)} - E_{\overline{|1,0\rangle}}^{\rm (lf)}\, ,
    \label{omega-target-01}\ee
where $E_{\overline{|n,m\rangle}}^{\rm (lf)}$ is the laboratory-frame eigenenergy of the state $\overline{|n,m\rangle}$. We will call  ``$zz$-coupling'' the difference between these frequencies,
    \be
\omega_{zz}\equiv \omega_{\rm t}^{\rm c1} - \omega_{\rm t}^{\rm c0}= E_{\overline{|11\rangle}}+ E_{\overline{|00\rangle}}- E_{\overline{|01\rangle}}-  E_{\overline{|10\rangle}},
    \label{zz-coupling}\ee
where this combination of eigenenergies is the same in the laboratory and rotating frames. The $zz$-coupling is mainly due to the repulsion of the energy level $|11\rangle$ from the levels $|02\rangle$ and $|20\rangle$, which gives the approximate value
    \be
    \omega_{zz} \approx \frac{2g^2}{\Delta+\eta_{\rm t}} -\frac{2g^2}{\Delta-\eta_{\rm c}} .
    \label{zz-coupling-an}   \ee
From Eq.\ (\ref{zz-coupling}) we see that the $zz$-coupling can be also defined as $\omega_{zz} = \omega_{\rm c}^{\rm t1} - \omega_{\rm c}^{\rm t0}$, where $\omega_{\rm c}^{\rm t0}$ and $\omega_{\rm c}^{\rm t1}$ are the eigenfrequencies of the control qubit for the target-qubit states $|0\rangle$ and $|1\rangle$, respectively. Nonzero $\omega_{zz}$ will be important for numerical results; however, it will be neglected for analytical and semi-analytical results in the next section; in particular, we will not distinguish between $\omega_{\rm t}^{\rm c0}$, $\omega_{\rm t}^{\rm c1}$, and $\omega_{\rm t}$.

\section{Analytical and semi-analytical analysis}
\label{sec-analytics}

\subsection{Ideal CR gate operation}
\label{sec-ideal}

There is no drive, $\varepsilon=0$, before and after the CR gate operation. Therefore, the initial and final two-qubit states should be considered in the eigenbasis $\overline{|n,m\rangle}$ of the Hamiltonian $H_{\rm qb}+H_g$. The drive Hamiltonian $H_\varepsilon$ couples these eigenstates, providing an evolution used in the CR gate.

As follows from Fig.\ \ref{fig3}, in the rotating frame based on the drive frequency $\omega_{\rm d}$, there is a near-resonance condition between states $|n,0\rangle$ and $|n,1\rangle$, which leads to a near-resonance between eigenstates $\overline{|n,0\rangle}$ and $\overline{|n,1\rangle}$, while other pairs of states are off-resonance. Therefore, as long as the perturbation produced by $H_\varepsilon$ is small enough, it effectively couples only states $\overline{|n,0\rangle}$ and $\overline{|n,1\rangle}$, and for the ideal effective Hamiltonian $H_{\rm CR}^{\rm ideal}$ of the CR gate we can write
    \begin{eqnarray}
&& H_{\rm CR}^{\rm ideal} - (H_{\rm qb} +H_g) =  \big( \tilde{\varepsilon}_0 \overline{|0,1\rangle } \,\, \overline{\langle 0,0|} + \tilde{\varepsilon}_1 \overline{|1,1\rangle } \,\, \overline{\langle 1,0|}
    \nonumber \\
&& \hspace{3.4cm} + \tilde{\varepsilon}_2 \overline{|2,1\rangle } \,\, \overline{\langle 2,0|} + ...\big) + {\rm h.c.},
    \label{Ham-ideal}\end{eqnarray}
where $\tilde{\varepsilon}_0$ is the amplitude of the {\it effective drive on the target qubit} when the control-qubit state is $|0\rangle$, $\tilde{\varepsilon}_1$ is the effective drive amplitude for the control-qubit state $|1\rangle$, etc. (for small $g$ there is almost no difference between the effective drive in the bare basis or eigenbasis). The effective drive amplitudes $\tilde\varepsilon_n$ depend on the actual drive amplitude $\varepsilon$ (in the linear approximation being proportional to $\varepsilon$).

Note that if we are interested only in the states $|0\rangle$ and $|1\rangle$ of the control qubit, then in the terminology  of Refs.\ \cite{Chow2011, Chow2012, Magesan2018, Ware2015} the  effective Hamiltonian (\ref{Ham-ideal}) can be written as
    \[
    \frac{\tilde{\varepsilon}_0-\tilde{\varepsilon}_1}{2}\, Z_{\rm c}X_{\rm t}+\frac{\tilde{\varepsilon}_0+\tilde\varepsilon_1}{2} \, I_{\rm c}X_{\rm t},
    \]
where the Pauli operators $Z_{\rm c}$ and $I_{\rm c}$ act on the control qubit and the operator $X_{\rm t}$ acts on the target qubit (here we assume real $\tilde\varepsilon_0$ and $\tilde\varepsilon_1$; otherwise, we also need $Y_{\rm t}$). The CNOT gate can be realized with this effective interaction by applying the drive pulse with duration $\tau_{\rm p}$, satisfying the condition
    \be
    \int_0^{\tau_{\rm p}} [2\tilde{\varepsilon}_1(t)-2\tilde{\varepsilon}_0(t)]\, dt =\pi \,\,\, ({\rm mod}\, 2\pi) ,
    \label{CNOT-basic}\ee
complemented with two one-qubit rotations. The additional $x$-rotation of the target qubit by the angle $-\int_0^{\tau_{\rm p}} 2 \tilde{\varepsilon}_0(t)\, dt$ compensates the target-qubit rotation for the  control-qubit state $|0\rangle$, also providing $x$-rotation by angle $\pi$ for the control-qubit state $|1\rangle$. Besides the $x$-rotation of the target qubit, the control qubit should be $z$-rotated by the angle $\pi/2$ (relative to the rotating frame of the control qubit). This is needed because the $x$-rotation of the target qubit by  angle $\pi$ produces the operation $-iX$ instead of the desired (for CNOT) operation $X$, thus requiring additional phase factor $i$ for the control-qubit state $|1\rangle$ (the same factor exists in a one-qubit $X$ gate, but it is not important since it is  an overall phase, in contrast to the phase difference in a controlled two-qubit operation). Note that the  factors of 2 in Eq.\ (\ref{CNOT-basic}) are needed because the Rabi frequency is twice larger than the drive matrix element in the Hamiltonian.

The effective drive amplitudes $\tilde\varepsilon_n$ in Eq.\ (\ref{Ham-ideal}) can be easily found (in the ideal lowest-order case) by comparing Eqs.\ (\ref{Ham}) and (\ref{Ham-ideal}), which gives
    \be
\tilde\varepsilon_n = \overline{\langle n,1|} H_\varepsilon \overline{|n,0\rangle} .
    \label{tilde-epsilon-n}\ee
To calculate $\tilde\varepsilon_n$ to the lowest order, let us assume $\delta =0$, i.e., the drive resonant with the bare target qubit (the difference between bare and eigenfrequencies is not important for these approximate calculations). Then using $\overline{| 0,0\rangle}=| 0,0\rangle$ (see Fig.\ \ref{fig3}) and the first-order approximation $\overline{| 0,1\rangle} = | 0,1\rangle-(g/\Delta) \, |1,0\rangle$ (normalization correction is of the second order), we find the linear approximation
    \be
    \tilde\varepsilon_0 =  -\frac{g}{\Delta}\, \varepsilon.
    \label{tilde-varepsilon-0}\ee
Similarly, using the first-order approximations $\overline{| 1,0\rangle} = | 1,0\rangle + (g/\Delta) \, |0,1\rangle$ and $\overline{| 1,1 \rangle} = | 1,1\rangle - [\sqrt{2}\,g/(\Delta-\eta_{\rm c})] \, |2,0\rangle + [\sqrt{3}\, g/(\Delta +\eta_{\rm t})]\, |0,2\rangle$  (see Fig.\ \ref{fig3}), we obtain approximation
   \be
    \tilde\varepsilon_1 =  \frac{g}{\Delta} \, \varepsilon - \frac{\sqrt{2}\, g}{\Delta-\eta_{\rm c}} \, \sqrt{2}\, \varepsilon = -\frac{g}{\Delta}\, \frac{\Delta+\eta_{\rm c}}{\Delta-\eta_{\rm c}}\, \varepsilon.
    \label{tilde-varepsilon-1}\ee
Also similarly, using approximations $\overline{| 2,0\rangle} = | 2,0\rangle + [\sqrt{2}\,g/(\Delta-\eta_{\rm c})] \, |1,1\rangle$ and $\overline{| 2,1 \rangle} = | 2,1\rangle - [\sqrt{3}\,g/(\Delta-2\eta_{\rm c})] \, |3,0\rangle + [2 g/(\Delta -\eta_{\rm c} +\eta_{\rm t})]\, |1,2\rangle$, we find
   \begin{eqnarray}
&&     \tilde\varepsilon_2 =  \frac{\sqrt{2}\, g}{\Delta-\eta_{\rm c}} \,\sqrt{2}\, \varepsilon - \frac{\sqrt{3}\, g}{\Delta-2\eta_{\rm c}} \, \sqrt{3}\, \varepsilon
\nonumber \\
&& \hspace{0.4cm}    = -\frac{g(\Delta+\eta_{\rm c})}{(\Delta-\eta_{\rm c})(\Delta-2\eta_{\rm c})}\, \varepsilon ,
    \label{tilde-varepsilon-2}
    \end{eqnarray}
and for an arbitrary control-qubit state $|n\rangle$, within the model (\ref{Ham})--(\ref{Ham-varepsilon}) we obtain
    \begin{eqnarray}
&& \tilde\varepsilon_n = \frac{ng\varepsilon}{\Delta-(n-1)\eta_{\rm c}} -\frac{(n+1)g\varepsilon}{\Delta-n\eta_{\rm c}}
    \nonumber \\
&& \hspace{0.5cm} =- \frac{g(\Delta +\eta_{\rm c})}{[\Delta-(n-1)\eta_{\rm c}] (\Delta-n\eta_{\rm c})} \, \varepsilon .
    \label{tilde-varepsilon-n-res}\end{eqnarray}

Since the target-qubit rotation for the control-qubit state $|0\rangle$ is usually compensated, most important are the differences of effective drive amplitudes from $\tilde\varepsilon_{0}$, e.g.,
    \begin{eqnarray}
&& \tilde\varepsilon_1- \tilde\varepsilon_0 = \frac{2g\eta_{\rm c}}{\Delta(\eta_{\rm c}-\Delta)}\, \varepsilon ,
    \label{tilde-varepsilon-1-0}\\
&&  \tilde\varepsilon_2- \tilde\varepsilon_0 = \frac{2g\eta_{\rm c}(\eta_{\rm c}-2\Delta)}{\Delta(\eta_{\rm c}-\Delta)(2\eta_{\rm c}-\Delta)}\, \varepsilon .
     \label{tilde-varepsilon-2-0}\end{eqnarray}

Note that these formulas depend on anharmonicity $\eta_{\rm c}$ of the control qubit but do not depend on the target-qubit anharmonicity $\eta_{\rm t}$. Also, for $\eta_{\rm c}=0$ we have $\tilde\varepsilon_n=\tilde\varepsilon_0=-(g/\Delta)\varepsilon$. These properties are in agreement with the classical description of the CR gate operation discussed in Sec.\ \ref{Sec:System&Ham}.

\subsubsection*{Language of virtual-state transitions}
\label{sec-virtual}

Instead of using Eq.\ (\ref{tilde-epsilon-n}), we can find the effective drive amplitudes $\tilde\varepsilon_n$ (still to the first order) using the ideology of transitions via a virtual state. As seen in Fig.\ \ref{fig3}, it is possible to go from the state $|0,0\rangle$ to the resonant state $|0,1\rangle$ in two jumps: $|0,0\rangle\to |1,0\rangle \to |0,1\rangle$, which have transition amplitudes (matrix elements) $\varepsilon$ and $g$,  with the intermediate state separated by the energy difference $\Delta$. Therefore, the amplitude of this transition (effective coupling between states $|0,0\rangle$ and $|0,1\rangle$) is
   \be
    \tilde\varepsilon_0 =  \varepsilon \, \frac{-1}{\Delta} \, g ,
    \ee
which coincides with Eq.\ (\ref{tilde-varepsilon-0}).

For the transition between states $|1,0\rangle$ and $|1,1\rangle$, there are  two two-jump paths: via the state $|2,0\rangle$ (which is higher in energy by $\Delta-\eta_{\rm c}$) and via $|0,1\rangle$ (which is lower in energy by $\Delta$). Adding these two amplitudes, we obtain
   \be
    \tilde\varepsilon_1 =  \sqrt{2}\,\varepsilon \,   \frac{-1}{\Delta-\eta_{\rm c}} \, \sqrt{2}\, g  + g\,  \frac{-1}{-\Delta} \, \varepsilon  ,
    \ee
which coincides with Eq.\ (\ref{tilde-varepsilon-1}).

Similarly, adding the amplitudes for the paths $|n,0\rangle \to |n+1,0\rangle \to |n,1\rangle$ and $|n,0\rangle \to |n-1,1\rangle \to |n,1\rangle$, we obtain
   \be
    \tilde\varepsilon_n =  - \frac{\sqrt{n+1}\,\varepsilon\, \sqrt{n+1}\, g}
    { \Delta-n\eta_{\rm c}} + \frac{\sqrt{n}\, g\, \sqrt{n}\, \varepsilon}{\Delta-(n-1)\eta_{\rm c}},
    \ee
which coincides with Eq.\ (\ref{tilde-varepsilon-n-res}).

\subsection{Next-order analytics}
\label{sec-next-order}

Numerical results for the effective drive amplitudes $\tilde\varepsilon_n$ (discussed later) show that $\tilde\varepsilon_n$ is proportional to the actual drive amplitude $\varepsilon$ [as expected from Eq.\ (\ref{tilde-varepsilon-n-res})] only in some range of $\varepsilon$-values. A minor deviation from the linearity at very small $\varepsilon$ (discussed  later) is due to dependence of the target-qubit frequency on the control-qubit state -- see Eq.\ (\ref{zz-coupling}). The deviation from linearity at large $\varepsilon$ is much more important for practice since it makes it impossible to shorten the CNOT gate duration beyond some value by simply increasing the drive amplitude.

In order to understand the reason for the deviation from linearity at large $\varepsilon$, in this section we develop the next-order analytics for $\tilde\varepsilon_0$ and $\tilde\varepsilon_1$, which gives corrections compared with Eqs.\ (\ref{tilde-varepsilon-0}) and (\ref{tilde-varepsilon-1}). Note that a similar next-order analytics has been developed in Ref.\ \cite{Magesan2018}, though in a very different way (after a misprint correction, the result of Ref.\ \cite{Magesan2018} coincides with our result).

The simple analytics (\ref{tilde-varepsilon-0})--(\ref{tilde-varepsilon-n-res}) has been obtained from Eq.\ (\ref{tilde-epsilon-n}), which treats the drive Hamiltonian $H_\varepsilon$ as a small perturbation. However, for a large drive amplitude $\varepsilon$, the eigenbasis of $H_{\rm qb}+H_g$ is no longer the appropriate eigenbasis; instead,  $H_{\rm qb}+H_\varepsilon$ is the main Hamiltonian, while $H_g$ is the perturbation.
Note that in the linear approximation, the same $\tilde\varepsilon_{n}$ as in Eq.\ (\ref{tilde-epsilon-n}) can be obtained by exchanging the roles of $H_g$ and $H_\varepsilon$, i.e., by using
    \be
\tilde\varepsilon_n = \overline{_\varepsilon \! \langle n,1|} H_g \overline{|n,0\rangle_\varepsilon},
    \label{tilde-epsilon-n-2}\ee
where $\overline{|n,m\rangle_\varepsilon}$ denotes the eigenstate of $H_{\rm qb}+H_\varepsilon$. This equivalence is clear from the discussed above approach of virtual-state transitions, which treats $H_g$ and $H_\varepsilon$ on equal footing.

For a large $\varepsilon$, Eq.\ (\ref{tilde-epsilon-n-2}) is more appropriate than Eq.\ (\ref{tilde-epsilon-n}) to calculate $\tilde\varepsilon_n$. Even though the initial and final states should still be treated in the eigenbasis of $H_{\rm qb}+H_g$, during the front and rear ramps of the microwave pulse  the appropriate eigenbases essentially transform into each other, leading to Eq.\ (\ref{tilde-epsilon-n-2}). While we do not have a rigorous justification of the approximation (\ref{tilde-epsilon-n-2}) (only a general understanding in the spirit of the adiabatic theorem), numerical results confirm its good accuracy.

Since for the eigenstates $\overline{|n,m\rangle_\varepsilon}$ used in Eq.\ (\ref{tilde-epsilon-n-2}) the ladders in Fig.\ \ref{fig3} are uncoupled, we can write
    \be
\overline{|n,0\rangle_\varepsilon} =\overline{|n\rangle_\varepsilon}\, |0\rangle_{\rm t} , \,\,\, \overline{|n,1\rangle_\varepsilon} =\overline{|n\rangle_\varepsilon}\, |1\rangle_{\rm t} ,
    \ee
where $\overline{|n\rangle_\varepsilon}$ are the control-qubit eigenstates, which account for the drive. They satisfy the Schr\"odinger equation
    \be
H_{\rm qb +\varepsilon}^{\rm (c)}\, \overline{|n\rangle_\varepsilon} = E_{\,\overline{|n\rangle_\varepsilon}} \, \overline{|n\rangle_\varepsilon}
    \label{Schrod-eq-control}\ee
with the Hamiltonian for only the control qubit,
    \be
H_{\rm qb +\varepsilon}^{\rm (c)}= \sum_n E_n^{\rm (c)}|n\rangle \langle n| + \sqrt{n}\, (\varepsilon \, |n\rangle \langle n-1|+\varepsilon^* |n-1\rangle\langle n|) .
    \label{Ham-control-qubit+drive}\ee
Then solving this Schr\"odinger equation and finding the eigenstates,
    \be
 \overline{|n\rangle_\varepsilon} = \sum\nolimits_k  c_k^{(n)} |k\rangle ,
    \label{n-eigen-control}\ee
we find the effective drive amplitudes $\tilde\varepsilon_n$ from Eq.\ (\ref{tilde-epsilon-n-2}) as   (see Fig.\ \ref{fig3})
    \be
    \tilde\varepsilon_n = g \sum\nolimits_k \sqrt{k}\,  c_k^{(n)} \left( c_{k-1}^{(n)} \right)^* .
    \label{tilde-epsilon-n-1q}\ee

Let us use this approach to find $\tilde\varepsilon_0$ up to the order $\varepsilon^3$ [instead of  $\varepsilon^1$ in the linear approximation (\ref{tilde-varepsilon-0})], treating $H_\varepsilon$ as a perturbation of $H_{\rm qb}$. The eigenstate $\overline{|0\rangle_\varepsilon}$ of the control qubit can be written as
    \be
\overline{|0\rangle_\varepsilon}=\frac{|0\rangle+\alpha \, |1\rangle +\beta \, |2\rangle +\gamma \, |3\rangle + ...}{\cal N} ,
    \label{0-eps-gen}\ee
where $\cal N$ is a normalization. Substituting this form into the Schr\"odinger equation (\ref{Schrod-eq-control}) and equating the coefficients for the basis states $|0\rangle$, $|1\rangle$, and $|2\rangle$, we obtain
    \begin{eqnarray}
&&   \varepsilon \alpha  =E,
   \label{contrib-0} \\
&& E_1 \alpha  + \sqrt{2}\, \varepsilon \, \beta +\varepsilon = E\, \alpha ,
    \label{contrib-1}\\
&& E_2\beta + \sqrt{2}\, \varepsilon \, \alpha + \sqrt{3}\, \varepsilon \, \gamma = E\, \beta ,
    \label{contrib-2}\end{eqnarray}
where for brevity $E=E_{\overline{|0\rangle_\varepsilon}}$, $E_n = E_n^{\rm (c)}$, we used $E_0=0$ [as in Eq.\ (\ref{E-c-n})] and also assumed that $\varepsilon$ is real.

To the lowest order, assuming small $\varepsilon$ (therefore $\gamma \ll \beta \ll \alpha$ and $E\approx 0$) we crudely find
    \be
    \alpha \approx \frac{-\varepsilon}{E_1}, \,\,\, \beta \approx \frac{-\sqrt{2}\, \varepsilon\, \alpha}{E_2}\approx \frac{\sqrt{2}\,\varepsilon^2}{E_1 E_2}, \,\,\, E \approx \frac{-\varepsilon^2}{E_1} .
    \label{alpha-beta-crude}\ee
Using these values for $\beta$ and $E$ in Eq.\ (\ref{contrib-1}), we obtain a better approximation (up to $\varepsilon^3$) for $\alpha$:
    \be
\alpha \approx -\frac{\varepsilon (1+2\varepsilon^2/E_1E_2)}{E_1+\varepsilon^2/E_1} \approx -\frac{\varepsilon}{E_1}\bigg( 1+\frac{2\varepsilon^2}{E_1E_2} -\frac{\varepsilon^2}{E_1^2}\bigg) .
    \label{alpha-better}\ee

To find $\tilde\varepsilon_0$ with accuracy up to $\varepsilon^3$, we need $\alpha$ with accuracy up to $\varepsilon^3$, $\beta$ with accuracy up to $\varepsilon^2$ and $\cal N$ with accuracy up to $\varepsilon^2$, while $\gamma$ is not needed [see Eq.\ (\ref{tilde-epsilon-n-1q})]. Thus, we use Eq.\ (\ref{alpha-better}) for $\alpha$, Eq.\ (\ref{alpha-beta-crude}) for $\beta$, and ${\cal N}\approx 1+ (\varepsilon/E_1)^2/2$ to obtain
    \begin{eqnarray}
&& \overline{|0\rangle_\varepsilon} \approx \bigg( 1-\frac{\varepsilon^2}{2E_1^2}\bigg) |0\rangle -\frac{\varepsilon}{E_1} \bigg( 1+\frac{2\varepsilon^2}{E_1E_2} -\frac{3\varepsilon^2}{2E_1^2}\bigg)  |1\rangle
    \nonumber \\
&& \hspace{1cm} + \frac{\sqrt{2}\, \varepsilon^2}{E_1 E_2}\, |2\rangle .
    \label{eigen-0-epsilon}\end{eqnarray}
The energy of state $\overline{|0\rangle_\varepsilon}$ (not needed for this derivation but needed later) is
    \be
E_{\,\overline{|0\rangle_\varepsilon}} = \varepsilon \alpha \approx
-\frac{\varepsilon^2}{E_1} \bigg( 1+\frac{2\varepsilon^2}{E_1E_2} -\frac{\varepsilon^2}{E_1^2} \bigg) .
    \label{eigenenergy-0}\ee
Finally, using Eqs.\ (\ref{tilde-epsilon-n-1q}) and (\ref{eigen-0-epsilon}), we obtain
    \be
\tilde\varepsilon_0 =  -g \, \frac{\varepsilon}{E_1} \bigg( 1-\frac{2\varepsilon^2}{E_1^2}+ \frac{4\varepsilon^2}{E_1 E_2} \bigg)
    \label{tilde-varepsilon-0-next}\ee
with accuracy up to $\varepsilon^3$. Note that $E_1=E_1^{\rm (c)}=\Delta+\delta$,  $E_2=E_2^{\rm (c)}=2(\Delta+\delta)-\eta_{\rm c}$, and we can neglect $\delta$ (i.e., use $\delta=0$). For a complex $\varepsilon$, we need to replace $\varepsilon^2$ in parentheses with $|\varepsilon|^2$.

Calculation of $\tilde\varepsilon_1$ up to the order $\varepsilon^3$ is similar and requires finding $\overline{|1\rangle_\varepsilon}$.  Note that the calculations are easier if the energies are counted  from $E_1$, because then the eigenenergy $E$ in equations similar to Eqs.\ (\ref{contrib-0})--(\ref{contrib-2}) is small. The calculations give
    \begin{eqnarray}
&&\hspace{-0.3cm}\overline{|1\rangle_\varepsilon} \approx \bigg( 1-\frac{\varepsilon^2}{2E_{01}^2} -\frac{\varepsilon^2}{E_{21}^2}\bigg) |1\rangle
    \nonumber \\
&& - \frac{\sqrt{2}\, \varepsilon}{E_{21}}\bigg( 1+\frac{3\varepsilon^2}{E_{21}E_{31}} - \frac{\varepsilon^2}{E_{01}E_{21}}-\frac{3\varepsilon^2}{E_{21}^2} -\frac{\varepsilon^2}{2E_{01}^2}\bigg)  |2\rangle
      \nonumber \\
&&  -\frac{\varepsilon}{E_{01}} \bigg( 1-\frac{3\varepsilon^2}{2E_{01}^2}-\frac{2\varepsilon^2}{E_{01}E_{21}} -\frac{\varepsilon^2}{E_{21}^2}\bigg) |0\rangle
 +\frac{\sqrt{6}\,\varepsilon^2}{E_{21}E_{31}} |3\rangle   ,
    \nonumber \\
    \label{eigen-1-epsilon}
    \end{eqnarray}
where $E_{nn'}\equiv E_n-E_{n'}=E_n^{\rm (c)}-E_{n'}^{\rm (c)}$. The corresponding energy is
    \begin{eqnarray}
&& E_{\,\overline{|1\rangle_\varepsilon}} \approx E_1 -\frac{\varepsilon^2}{E_{01}} \bigg( 1- \frac{\varepsilon^2}{E_{01}^2} -\frac{2\varepsilon^2}{E_{21}E_{01}} \bigg)
    \nonumber \\
&& \hspace{1.1cm} - \frac{2\varepsilon^2}{E_{21}} \bigg( 1+\frac{3\varepsilon^2}{E_{21}E_{31}} -\frac{2\varepsilon^2}{E_{21}^2} -\frac{\varepsilon^2}{E_{01}E_{21}} \bigg) . \qquad
    \label{eigenenergy-1}\end{eqnarray}
Using Eqs.\ (\ref{tilde-epsilon-n-1q}) and (\ref{eigen-1-epsilon}), we obtain
    \begin{eqnarray}
&& \tilde\varepsilon_1 = -\frac{2\varepsilon g}{E_{21}} \bigg(  1+\frac{6\varepsilon^2}{E_{21}E_{31}}+ \frac{\varepsilon^2}{E_{10}E_{21}}-\frac{4\varepsilon^2}{E_{21}^2} -\frac{\varepsilon^2}{E_{10}^2}\bigg)
    \nonumber \\
&& \hspace{0.8cm} +\frac{\varepsilon g}{E_{10}} \bigg( 1-\frac{2\varepsilon^2}{E_{10}^2}+\frac{2\varepsilon^2}{E_{10}E_{21}} -\frac{2\varepsilon^2}{E_{21}^2}\bigg) \qquad
    \label{tilde-varepsilon-1-next}\end{eqnarray}
with accuracy up to $\varepsilon^3$ (note the use of $E_{10}$ instead of $E_{01}$ in the preceding formulas). In this formula $E_{10}=\Delta+\delta$, $E_{21}=\Delta+\delta-\eta_{\rm c}$, $E_{31}=2(\Delta+\delta)-3\eta_{\rm c}$, and we can neglect $\delta$ (i.e., $\delta=0$). For a complex $\varepsilon$, we need to replace $\varepsilon^2$ in parentheses with $|\varepsilon|^2$.

\begin{figure}[t]
\includegraphics[width=0.95\linewidth]{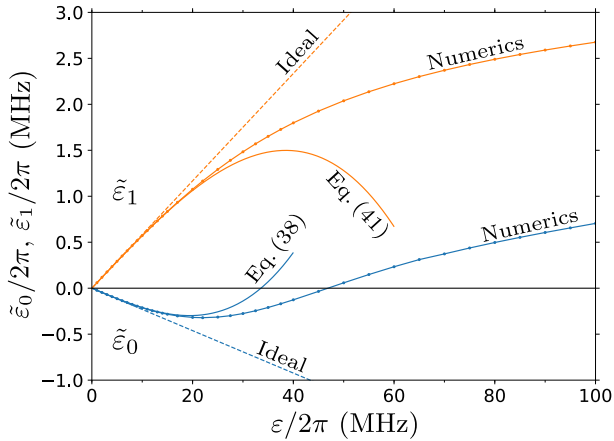}
	\caption{The effective drive amplitudes $\tilde\varepsilon_0$ (blue lines) and $\tilde\varepsilon_1$ (orange lines) as functions of the drive amplitude $\varepsilon$, calculated using the ideal-case approximation, Eqs.\ (\ref{tilde-varepsilon-0}) and (\ref{tilde-varepsilon-1}) (dashed straight lines), the third-order formulas (\ref{tilde-varepsilon-0-next}) and (\ref{tilde-varepsilon-1-next}) (solid lines without symbols), and numerically (solid lines with symbols).  We used the qubit-qubit coupling $g/2\pi = 3$ MHz, qubit anharmonicity $\eta_{\rm c}/2\pi = \eta_{\rm t}/2\pi = 300$ MHz, and detuning $\Delta/2\pi=130$ MHz.
 }
	\label{fig-next-order}
\end{figure}

Figure \ref{fig-next-order} shows the effective drive amplitudes $\tilde\varepsilon_0$ and $\tilde\varepsilon_1$ as functions of the actual drive amplitude $\varepsilon$ calculated in several ways for the following parameters (which are some typical experimental parameters): $g/2\pi = 3$ MHz, $\eta_{\rm c}/2\pi=300$  MHz, $\Delta/2\pi = 130$ MHz. The blue lines (which initially go down) show $\tilde\varepsilon_0$, the orange lines (which initially go up) show $\tilde\varepsilon_1$. The solid lines without symbols are calculated using Eqs.\ (\ref{tilde-varepsilon-0-next}) and (\ref{tilde-varepsilon-1-next}) (using $\delta=0$), while the straight dashed lines represent the simple linear approximation, Eqs.\ (\ref{tilde-varepsilon-0}) and (\ref{tilde-varepsilon-1}). The solid lines with symbols show the numerical results (the numerical procedure is described later in Sec.\ \ref{sec-numerical}, for numerics we assume $\eta_{\rm t}=\eta_{\rm c}$ and  $\omega_{\rm d}=\omega_{\rm t}^{\rm c0}$).

We see that the third-order approximation [Eqs.\ (\ref{tilde-varepsilon-0-next}) and (\ref{tilde-varepsilon-1-next})] correctly describes the deviation of the dependences $\tilde\varepsilon_0(\varepsilon)$ and $\tilde\varepsilon_1(\varepsilon)$ from ideal straight lines at relatively small $\varepsilon$, but fails to fit well the case of relatively large $\varepsilon$. This is because higher-order terms become important even for moderate values of $\varepsilon$. The problem has a similarity with the poor performance of the perturbation approach in the analysis of the circuit QED measurement of transmons when the number of photons is comparable to the critical number.

\begin{figure}[t]
\includegraphics[width=0.95\linewidth]{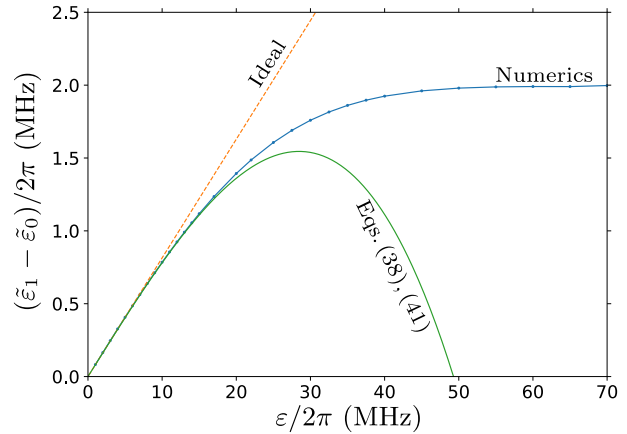}
	\caption{The CR gate speed $\tilde\varepsilon_1 -\tilde\varepsilon_0$ as a function of the drive amplitude $\varepsilon$, calculated using the linear-order approximation (\ref{tilde-varepsilon-1-0}) (dashed straight line), third-order approximations (\ref{tilde-varepsilon-0-next}) and (\ref{tilde-varepsilon-1-next}) (green solid line), and numerically (blue solid line with symbols). The parameters are the same as in Fig.\ \ref{fig-next-order}.  }
	\label{fig-our-IBM}
\end{figure}

Figure \ref{fig-our-IBM} shows the difference $\tilde\varepsilon_1-\tilde\varepsilon_0$ as a function of $\varepsilon$ (for the same parameters as in Fig.\ \ref{fig-next-order}), also calculated in several ways. The dashed straight line corresponds to the simple formula (\ref{tilde-varepsilon-1-0})  for the ideal operation. The numerical results are shown by the blue solid line with symbols. The green solid line is calculated using Eqs.\  (\ref{tilde-varepsilon-0-next}) and (\ref{tilde-varepsilon-1-next}). We have checked (analytically and numerically) that this line coincides with the result given by Eq.\ (4.25) of Ref.\ \cite{Magesan2018} (after correction of a misprint in the initial version of Ref.\ \cite{Magesan2018}; the translation of notations is $g=J_{\text{\cite{Magesan2018}}}$, $2\varepsilon =\Omega_{\text{\cite{Magesan2018}}}$, $\Delta =\Delta_{\text{\cite{Magesan2018}}}$, $\eta_{\rm c}=-\delta_{1\, \text{\cite{Magesan2018}}}$, $\tilde\varepsilon_0 - \tilde\varepsilon_1 = (ZX/2)_{{\rm coeff}\,\text{\cite{Magesan2018}}}$).
Most importantly, from Fig.\ \ref{fig-our-IBM} we see that the third-order approximation correctly describes initial deviation of the CR gate speed from the linear-order result, but cannot be used for quantitative analysis in the practically interesting regime of large drive amplitudes.

\subsection{Semi-analytical results}
\label{sec-semi-analytics}

The method developed in the previous section can be naturally extended to arbitrary large drive amplitudes $\varepsilon$. For that the eigenstates $\overline{|n\rangle_\varepsilon}$ of the control-qubit Hamiltonian (\ref{Ham-control-qubit+drive}) can be found numerically, and then the effective drive amplitudes $\tilde\varepsilon_n$ can be calculated using Eq.\ (\ref{tilde-epsilon-n-1q}). Since numerical diagonalization of a Hamiltonian for few levels is very easy (compared with full numerical simulation of the two-qubit evolution discussed in the next section), we call this method semi-analytical.

\begin{figure}[t]
\includegraphics[width=0.95\linewidth]{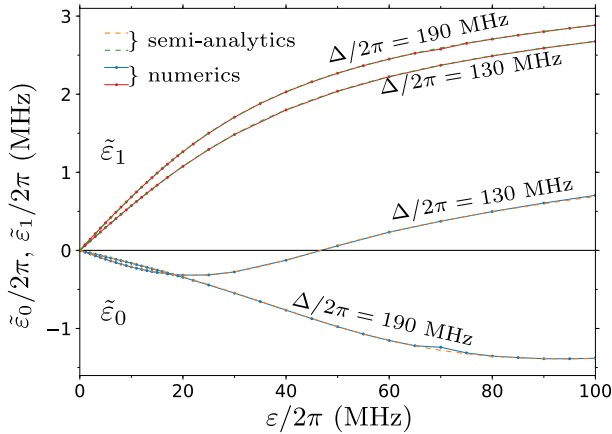}
	\caption{Effective drive amplitudes $\tilde\varepsilon_0$ and $\tilde\varepsilon_1$ as functions of $\varepsilon$, calculated numerically (solid lines with symbols) and using the semi-analytical approach, Eqs.\ (\ref{Schrod-eq-control})--(\ref{tilde-epsilon-n-1q}) (dashed lines, practically coinciding with the solid lines). We used $g/2\pi = 3$ MHz, $\eta_{\rm c}/2\pi=\eta_{\rm t}/2\pi  = 300$ MHz, and two values for the detuning:  $\Delta/2\pi=130$ MHz and 190 MHz.  }
	\label{fig-semi-analytics-comp}
\end{figure}

Figure \ref{fig-semi-analytics-comp} shows a comparison of the semi-analytical results (dashed lines) for $\tilde\varepsilon_0$ and $\tilde\varepsilon_1$ with the numerical results (solid lines with symbols, the numerical procedure is discussed in Sec.\ \ref{sec-numerical}). The parameters are the same as in Figs.\ \ref{fig-next-order} and \ref{fig-our-IBM}, except we use two values of the detuning: $\Delta/2\pi= 130$ MHz and 190 MHz. In the semi-analytics, we use 7 levels of the control qubit. We see that the  numerical results agree with semi-analytics very well for all values of the drive amplitude $\varepsilon$ (the lines are practically indistinguishable, except for the lowest lines at around 70 MHz, where a minor difference is caused by a resonance between levels  $\overline{|0,1\rangle_\varepsilon}$ and  $\overline{|1,2\rangle_\varepsilon}\,$). Similarly, we found a very good agreement for other values of the parameters as well. Therefore, the semi-analytical method based on Eqs.\ (\ref{Schrod-eq-control})--(\ref{tilde-epsilon-n-1q}) seems to be a sufficiently simple and accurate way of analyzing the dependence of the CR gate speed on parameters.

Note that $\tilde\varepsilon_n$ in the semi-analytical method is proportional to the qubit-qubit coupling $g$ and  also depends on two dimensionless ratios: $\Delta/\eta_{\rm c}$ and $\varepsilon/\eta_{\rm c}$  (assuming $\delta=0$). In the Duffing (Kerr) approximation  (\ref{E-c-n}), these two ratios fully define the eigenstates (\ref{n-eigen-control}) (in a better approximation \cite{Khezri2018, Sank2016} the results would also depend on the dimensionless parameter $\eta_{\rm c}/\omega_{\rm c}$). Therefore, in our analysis the ratio $\tilde\varepsilon_n/g$ is a function of {\it only two parameters}: $\Delta/\eta_{\rm c}$ and $\varepsilon/\eta_{\rm c}$.

\begin{figure}[t]
\includegraphics[width=\linewidth]{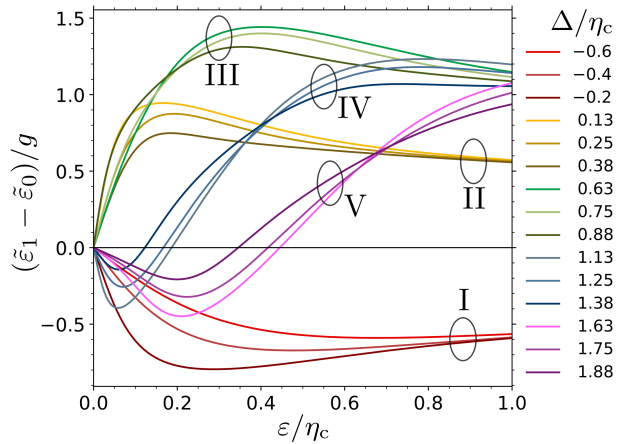}
	\caption{Dimensionless CR gate speed $(\tilde{\varepsilon}_1 -\tilde{\varepsilon}_0)/g$ as a function of the dimensionless drive amplitude $\varepsilon /\eta_{\rm c}$ for several values of the dimensionless detuning $\Delta/\eta_{\rm c}$. The lines are calculated using the semi-analytical method (\ref{Schrod-eq-control})--(\ref{tilde-epsilon-n-1q}). At large $\varepsilon$, the lines  group into ``bands''. The group I is for $\Delta/\eta_{\rm c}<0$, the group II is for $\Delta/\eta_{\rm c}$ in the interval $(0,\, 1/2)$. Similarly, the groups III, IV and V are for $\Delta/\eta_{\rm c}$ in the intervals $(1/2,\, 1)$, $(1,\, 3/2)$, and $(3/2,\, 2)$, respectively.
 }
	\label{fig-speed-semi-analyt}
\end{figure}

Figure \ref{fig-speed-semi-analyt} shows the dimensionless speed $(\tilde\varepsilon_1-\tilde\varepsilon_0)/g$ of the CR gate as a function of the dimensionless drive amplitude $\varepsilon/\eta_{\rm c}$ for several values of the dimensionless detuning $\Delta/\eta_{\rm c}$ (the lines are calculated using the semi-analytical method). While the behavior at small $\varepsilon$ agrees with Eq.\ (\ref{tilde-varepsilon-1-0}) (not shown), the behavior at large $\varepsilon$ mostly depends on whether the detuning $\Delta=\omega_{\rm c}-\omega_{\rm t}$ is negative or positive and on the integer part of the ratio $2\Delta/\eta_{\rm c}$ for positive $\Delta$. As seen in Fig.\ \ref{fig-speed-semi-analyt}, at large $\varepsilon$ the lines group according to the interval to which $\Delta$ belongs: $(-\infty, 0)$, $(0,\eta_{\rm c}/2)$, $(\eta_{\rm c}/2, \eta_{\rm c})$, $(\eta_{\rm c}, 3\eta_{\rm c}/2)$, $(3\eta_{\rm c}/2, 2\eta_{\rm c})$, etc.\ (in Fig.\ \ref{fig-speed-semi-analyt} these groups of lines are labeled sequentially as I, II, III, etc.). We do not show the lines for $\Delta/\eta_{\rm c}=0$, 1/2, 1, 3/2, etc.\ because at these values there is a resonance between the levels, $E^{\rm (c)}_n=E^{\rm (c)}_0$ and $E^{\rm (c)}_{n-1}=E^{\rm (c)}_1$  for $n=2\Delta/\eta_{\rm c}+1$ [see Eq.\ (\ref{E-c-n}) for $\delta =0$], and correspondingly the CR gate does not operate as intended (due to a very large leakage -- see below), also leading to computational problems in the semi-analytical and numerical calculations.

It is simple to understand why the lines in Fig.\ \ref{fig-speed-semi-analyt} group into ``bands''  at large $\varepsilon$. In the solution of the Schr\"odinger equation (\ref{Schrod-eq-control}) for the Hamiltonian (\ref{Ham-control-qubit+drive}) at large $\varepsilon$, the main effect is a strong level repulsion, which depends on the relative position (topology) of the bare energy levels $E_n^{\rm (c)}$ (i.e., which level is in between which levels; this topology does not change with $\varepsilon$ because of the adiabatic theorem). In contrast, the level repulsion does not depend much on a particular value of the initial bare level difference, since the effect of $\varepsilon$ dominates. Therefore, at very large $\varepsilon$ the eigenstates (\ref{n-eigen-control}) do not depend on a particular value of $\Delta/\eta_{\rm c}$, but only on the integer part of $2\Delta/\eta_{\rm c}$ (for $\Delta >0$), which defines the topological structure of the ladder $E_n^{\rm (c)}$ (as mentioned above, the bare levels $E_n^{\rm (c)}$  intersect at integer values of  $2\Delta/\eta_{\rm c}$). Consequently, at very large $\varepsilon$ the effective drive amplitudes (\ref{tilde-epsilon-n-1q})  depend only on the integer part of $2\Delta/\eta_{\rm c}$. This is why there is a grouping of lines in Fig.\ \ref{fig-speed-semi-analyt}.

We see that most of the lines in Fig.\ \ref{fig-speed-semi-analyt} (all the lines in the experimentally important groups II and III) have a maximum, with a relatively minor decrease after it (experimental results \cite{Chow2011, Ware2015} are somewhat similar). Experimentally, faster speed $(\tilde\varepsilon_1-\tilde\varepsilon_0)$ means shorter CNOT gate duration and therefore there is no benefit to increase the drive amplitude beyond the maximum in Fig.\ \ref{fig-speed-semi-analyt}.

\begin{figure}[t]
\includegraphics[width=0.95\linewidth]{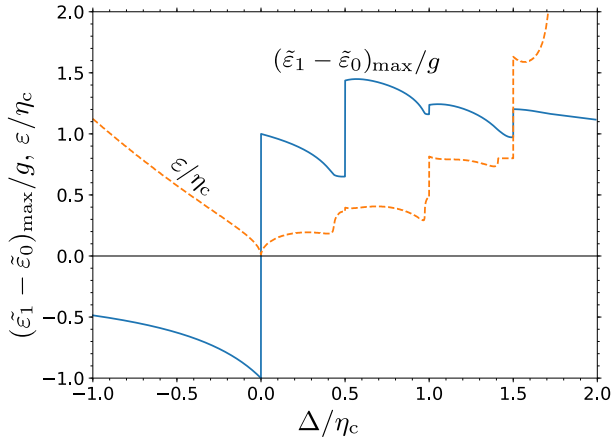}
	\caption{Maximized (minimized for negative values) dimensionless speed $(\tilde{\varepsilon}_1 -\tilde{\varepsilon}_0)_{\rm max}/g$ (solid blue line) and corresponding dimensionless drive amplitude $\varepsilon /\eta_{\rm c}$
(dashed orange line), as functions of the dimensionless detuning $\Delta/\eta_{\rm c}$. The lines are calculated using the semi-analytical method (\ref{Schrod-eq-control})--(\ref{tilde-epsilon-n-1q}).
 }
	\label{fig-speed-eps-max}
\end{figure}

The solid blue line in Fig.\ \ref{fig-speed-eps-max} shows the maximum value $(\tilde{\varepsilon}_1 -\tilde{\varepsilon}_0)_{\rm max}/g$ of the dimensionless speed  (or the minimum value for the negative speed) as a function of the dimensionless detuning $\Delta/\eta_{\rm c}$. The dimensionless drive amplitude $\varepsilon/\eta_{\rm c}$ at which this maximum is reached is shown by the dashed orange line. We see that the maximum CR gate speed is reached for the detuning $\Delta$ between $\eta_{\rm c}/2$ and $\eta_{\rm c}$ (group III in Fig.\ \ref{fig-speed-semi-analyt}). Note that our semi-analytical approach cannot be applied in close vicinities of the detunings $\Delta/\eta_{\rm c}=0$, 0.5, 1, 1.5, etc. The dependence of $(\tilde{\varepsilon}_1 -\tilde{\varepsilon}_0)_{\rm max}$ on $\Delta$ was measured experimentally \cite{Ware2015} and it showed a crudely similar behavior.

\section{Numerical approach}
\label{sec-numerical}

Numerically, we simulate the quantum evolution due to the rotating-frame Hamiltonian (\ref{Ham})--(\ref{Ham-varepsilon}), taking into account 7 levels in the control qubit (the same number as in the semi-analytics) and 5 levels in the target qubit, so that there are 35 levels in total. The simulation is based on matrix exponentiation for a time-dependent Hamiltonian, using the second-order Magnus expansion \cite{Magnus1954}. We also tried to use the fourth-order Runge-Kutta method, but found that for our typical parameters it is almost an order of magnitude slower to reach the same desired accuracy.

We start with diagonalization of the time-independent part $H_{\rm qb}+H_g$ of the Hamiltonian, and then the whole simulation is done in the eigenbasis $\overline{|n,m\rangle}$ of $H_{\rm qb}+H_g$, with the time-dependent drive Hamiltonian $H_\varepsilon (t)$ (expanded in the eigenbasis) causing the evolution. In this way, we obtain a $35\times 35$  unitary evolution matrix $V$ (in the eigenbasis $\overline{|n,m\rangle}\,$) for a given pulse of the drive amplitude $\varepsilon(t)$ with duration $\tau_{\rm p}$ (the pulse shape is discussed later). The matrix $V$ is then projected onto the computational two-qubit subspace, thus producing a $4\times 4$ matrix $M$, which is no longer unitary (here projection means the simple elimination of all other elements). Note that the reduced matrix $M$ is still defined in the eigenbasis, consisting of states $\overline{|0,0\rangle}=|0,0\rangle$, $\overline{|0,1\rangle}$, $\overline{|1,0\rangle}$ and $\overline{|1,1\rangle}$.

To find {\it fidelity} of an operation, we compare the reduced matrix $M$ with the desired $4\times 4$ unitary operation, which we denote $U$. The fidelity between $M$ and $U$ is defined as \cite{Zanardi2004,Pedersen2007}
\begin{align}
	F_{MU} = \frac{\text{Tr}(M^{\dagger} M)}{d(d+1)} + \frac{|\text{Tr}(M^{\dagger}U)|^2}{d(d+1)},
\label{fidelity}
\end{align}
where $d=4$ is the dimension of the two-qubit Hilbert space. This definition of the gate fidelity is equal to the  final-state fidelity (squared overlap) averaged over all (pure) initial states in the two-qubit subspace; therefore, $F_{MU}$ corresponds to the fidelity in Randomized Benchmarking (assuming that the states leaked outside the computational subspace never come back).

Even though the final goal of the CR gate operation is to produce CNOT (after additional single-qubit rotations), in the numerical procedure the desired $U$ is obviously not the CNOT. Instead, for a given pulse $\varepsilon (t)$ (which produces some matrix $V$ and corresponding matrix $M$), we \textit{define} $U$ {\it as the closest two-qubit unitary} (i.e., which maximizes the fidelity $F_{MU}$), restricted to the following class:
\begin{align}
	U = e^{i\theta_0}|0\rangle\langle 0|_{\text{c}\,}e^{-i(\varphi_0/2)X_{\text{t}}} + e^{i\theta_1}|1\rangle\langle 1|_{\text{c}\,}e^{-i(\varphi_1/2)X_{\text{t}}} ,
\label{U-class}
\end{align}
where $|n\rangle\langle n|_\text{c}$ acts on the control qubit, while $X_\text{t}$ acts on the target qubit (as mentioned above, we use the eigenbasis of $H_{\rm qb}+H_g$ for both $M$ and $U$).
The condition (\ref{U-class}) means that the state of the control qubit does not change (in the eigenbasis). Also, for state $|0\rangle$ of the control qubit, the target qubit is rotated about $x$-axis by angle $\varphi_0$; similarly, for control-qubit state $|1\rangle$, the target qubit is rotated about $x$-axis by angle $\varphi_1$. Besides that, in Eq.\ (\ref{U-class})  there are phases $\theta_0$ and $\theta_1$; disregarding the unimportant overall phase, this can be interpreted as $z$-rotation of the control qubit by angle $\theta_1-\theta_0$. Without loss of generality, we could assume $\theta_0 = 0$ (while keeping the same $\theta_1-\theta_0$) since this affects only the overall phase, and the definition (\ref{fidelity}) of the fidelity $F_{MU}$ is insensitive to the overall phase of $U$. Note that Eqs.\ (\ref{fidelity}) and (\ref{U-class}) can be easily generalized to include the third state of the control qubit (to consider it as a qutrit); however, here we consider only the two-level subspace.

Thus, to find $U$ for a given pulse of $\varepsilon(t)$, we maximize $F_{MU}$ over parameters $\varphi_0$, $\varphi_1$, and $\theta_1-\theta_0$. Fortunately, these optimal angles are given by analytical formulas in terms of the matrix elements of $M$:
    \begin{eqnarray}
&& \varphi_0=-{\rm arg}  \bigg(  \frac{M_{11}+M_{22}+M_{12}+M_{21}}{M_{11}+M_{22}-M_{12}-M_{21}}\bigg),
    \\
&& \varphi_1=-{\rm arg}  \bigg(  \frac{M_{33}+M_{44}+M_{34}+M_{43}}{M_{33}+M_{44}-M_{34}-M_{43}}\bigg),
    \\
&& \theta_0= {\rm arg} [(M_{11}+M_{22})\cos (\varphi_0/2)
    \nonumber \\
&& \hspace{1.5cm} +i (M_{12}+M_{21})\sin(\varphi_0/2)],
    \\
&& \theta_1 = {\rm arg} [(M_{33}+M_{44})\cos (\varphi_1/2)
    \nonumber \\
&& \hspace{1.5cm} +i (M_{34}+M_{43})\sin(\varphi_1/2)],
    \end{eqnarray}
where the rows (and columns) 1, 2, 3, and 4 of the matrix $M$ correspond to the states $\overline{|00\rangle}$, $\overline{|01\rangle}$, $\overline{|10\rangle}$, and $\overline{|11\rangle}$, respectively.
In this way, for a given pulse $\varepsilon(t)$, the CR gate operation is characterized by 4 resulting parameters: the angles $\varphi_0$, $\varphi_1$, and $\theta_0-\theta_1$ of the unitary (\ref{U-class}) and also infidelity $1-F_{MU}$, which is due to leakage outside of the computational two-qubit subspace and also due to the computational-space unitary not fitting well the class (\ref{U-class}).

\begin{figure}[t]
\includegraphics[width=0.75\linewidth]{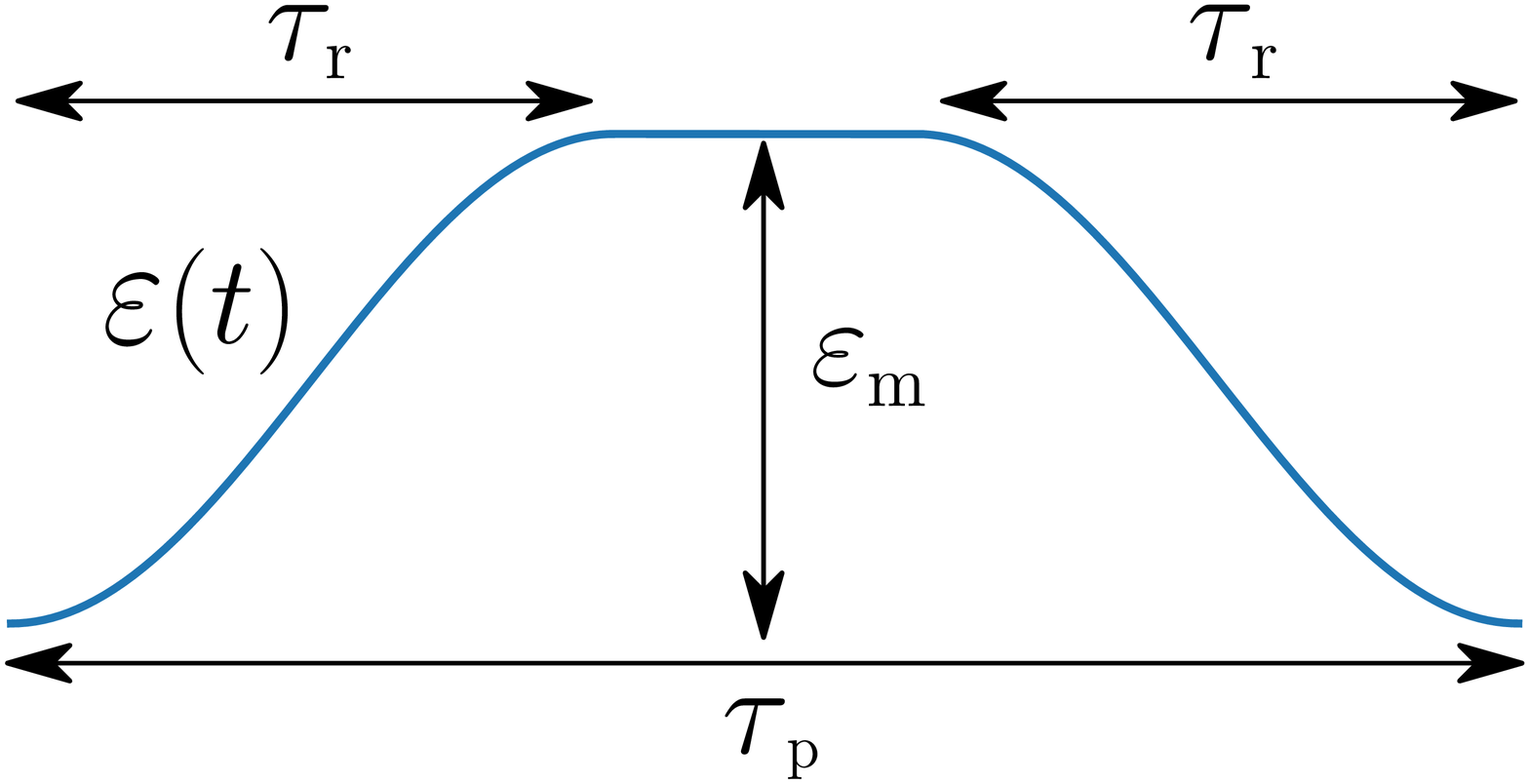}
	\caption{Pulse shape $\varepsilon(t)$ used in numerical simulations. The total pulse duration is $\tau_{\rm p}$, each cosine-shaped ramp has duration $\tau_{\rm r}$, the drive amplitude in the middle flat part is $\varepsilon_{\rm m}$.
}
	\label{fig-pulse-spape}
\end{figure}

We consider the pulse shape  $\varepsilon(t)$ of duration $\tau_{\rm p}$,
    \be
    \varepsilon(t) = \left\{
\begin{array}{ll}
      \displaystyle\frac{1-\cos (\pi t/\tau_{\rm r})}{2}\, \varepsilon_{\rm m},  & 0\leq t\leq \tau_{\rm r} ,
\vspace{0.2cm}      \\
      \varepsilon_{\rm m}, & \tau_{\rm r} \leq t\leq \tau_{\rm p}-\tau_{\rm r} ,
      \vspace{0.2cm} \\
      \displaystyle\frac{1-\cos [\pi (\tau_{\rm p}-t)/\tau_{\rm r}]}{2}\, \varepsilon_{\rm m},  & \tau_{\rm p}-\tau_{\rm r}\leq t\leq \tau_{\rm p} , \\
\end{array}
\right.
    \label{pulse-shape}\ee
which consists of the flat middle part with the {\it real} amplitude $\varepsilon_{\rm m}$ of the drive and two symmetric cosine-shaped ramps (so that there are no kinks), each with duration $\tau_{\rm r}$ -- see Fig.\ \ref{fig-pulse-spape}. As discussed later, sufficiently long ramps are needed to reduce leakage outside of the computational subspace.

Effective drive amplitudes $\tilde\varepsilon_0$ and $\tilde\varepsilon_1$ used in Sec.\ \ref{sec-analytics} (Figs.\ \ref{fig-next-order}--\ref{fig-semi-analytics-comp}) have been numerically calculated as the derivatives,
    \be
\tilde\varepsilon_0 = \frac{1}{2}\, \frac{\partial \varphi_0}{\partial \tau_{\rm p}}, \,\,\, \tilde\varepsilon_1 = \frac{1}{2} \, \frac{\partial \varphi_1}{\partial \tau_{\rm p}},
    \ee
while keeping the ramp duration $\tau_{\rm r}$, the middle-part amplitude $\varepsilon_{\rm m}$ (which replaces $\varepsilon$ in Sec.\ \ref{sec-analytics}), and other parameters fixed.

The small drive frequency detuning $\delta$ in numerical simulations is chosen in the following way. We first use the laboratory frame, i.e., $\delta =\omega_{\rm t}$ (for the Duffing oscillator model we can also use  $\delta=\omega_{\rm t}=0$) and calculate the eigenfrequencies of the target qubit $\omega_{\rm t}^{\rm c0}$ and $\omega_{\rm t}^{\rm c1}$ -- see Eq.\ (\ref{omega-target-01}).
If we want the drive to be exactly on-resonance with the target qubit when the control qubit is $|0\rangle$, then we need to use $\omega_{\rm d}=\omega_{\rm t}^{\rm c0}$, which gives $\delta = \omega_{\rm t}-\omega_{\rm t}^{\rm c0}$. Similarly, if we want $\omega_{\rm d}=\omega_{\rm t}^{\rm c1}$
(the drive on-resonance with the target qubit when the control qubit is $|1\rangle$), then we use $\delta = \omega_{\rm t}-\omega_{\rm t}^{\rm c1}$. If we want the drive frequency exactly in between the two resonances, $\omega_{\rm d}=(\omega_{\rm t}^{\rm c0}+\omega_{\rm t}^{\rm c1})/2$, then we use $\delta =\omega_{\rm t}-(\omega_{\rm t}^{\rm c0}+\omega_{\rm t}^{\rm c1})/2$.
Note that the frequency differences $\omega_{\rm t}-\omega_{\rm t}^{\rm c0}$ and
$\omega_{\rm t}-\omega_{\rm t}^{\rm c1}$ do not depend on a choice of the rotating frame.

Since experimentally the CR gate is mainly used to realize CNOT, in the numerical simulations we are mostly interested in the operations equivalent to CNOT up to single-qubit rotations. In analyzing the CNOT-equivalent gates, we usually use the pulse shape in which the
ramps occupy 30\% of the whole pulse duration each, i.e., $\tau_{\rm r}=0.3\,\tau_{\rm p}$ in Eq.\ (\ref{pulse-shape}) (Fig.\ \ref{fig-tau-ramp} is an exception). For a given middle-part amplitude $\varepsilon_{\rm m}$, we find the shortest pulse duration $\tau_{\rm p}$, for which
    \be
\varphi_1-\varphi_0 = \pi \,\,\, ({\rm mod}\, 2\pi).
    \ee
This is what we call the CNOT gate duration $\tau_{\rm p}^{\rm \scriptscriptstyle CNOT}(\varepsilon_{\rm m})$, neglecting durations of the additional single-qubit operations ($x$-rotation of the target qubit and $z$-rotation of the control qubit). We assume perfect fidelity of single-qubit operations; therefore, the CNOT infidelity is $1-F_{MU}$ for the pulse with duration $\tau_{\rm p}^{\rm \scriptscriptstyle CNOT}(\varepsilon_{\rm m})$.

In the simulations we fully neglect decoherence. A crude estimate of the fidelity decrease $\Delta F$ due to energy relaxation and pure dephasing can be obtained by considering idle qubits, which decohere during time $\tau_{\rm p}^{\rm \scriptscriptstyle CNOT}$. This gives the estimate
 \be
\Delta F \simeq \frac{1}{5}\, \frac{\tau_{\rm p}^{\rm \scriptscriptstyle CNOT}}{T_1^{\rm (c)}} + \frac{1}{5}\, \frac{\tau_{\rm p}^{\rm \scriptscriptstyle CNOT}}{T_1^{\rm (t)}} + \frac{2}{5}\, \frac{\tau_{\rm p}^{\rm \scriptscriptstyle CNOT}}{T_2^{\rm (c)}}+ \frac{2}{5}\, \frac{\tau_{\rm p}^{\rm \scriptscriptstyle CNOT}}{T_2^{\rm (t)}},
 \label{Delta-F}\ee
where $T_1^{\rm (c)}$ and $T_1^{\rm (t)}$  are the energy relaxation times for the control and target qubits, and similarly $T_2^{\rm (c)}$ and $T_2^{\rm (t)}$ are the dephasing times (which include contributions due to the energy relaxation and pure dephasing, $1/T_2=1/2T_1+1/T_\varphi$). This estimate is obtained by summing the single-qubit Pauli error rates $t/2T_1$ and $t/2T_\varphi$ and then converting the result into the 2-qubit average gate fidelity by using the factor $4/5$.
Note, however, that actual fidelity decrease $\Delta F$ can be significantly larger than the estimate (\ref{Delta-F}) because the CR gate operation involves a significant population of the level $|2\rangle$ and even higher levels of the control qubit, which have poorer coherence than the level $|1\rangle$.

One run of the evolution simulation for a given pulse duration typically takes a few seconds on a desktop or a laptop computer. Finding $\tau_\text{p}^{\rm \scriptscriptstyle CNOT}$ requires several tens of runs, so a typical time to produce a line showing dependence of the CNOT gate operation on $\varepsilon_\text{m}$ is few hours. The simulation time significantly depends on the number of time steps in the pulse ramps; we have used 600 time steps for each ramp, which gives a quite good accuracy for the simulations. For quick (and much less accurate) simulations it is possible to use $\sim$100 time steps per ramp.

Let us list the main approximations used in our numerics: (i) neglected decoherence, (ii) Duffing-oscillator approximation for the transmon energy levels, (iii) linear-oscillator approximation for the transmon matrix elements, (iv) direct coupling of qubits instead of coupling via resonator, (v) RWA, (vi) using only $7\times 5$ levels (we checked that this is sufficient), (vii) no microwave crosstalk (except in Fig.\ \ref{fig-crosstalk}), (viii) simple pulse shape without distortions, (ix) absence of neighboring qubits. In spite of a rather long list of approximations, we believe our simulation results give a reasonably accurate description of the intrinsic operation of the CR gate (neglecting decoherence, which in practice may give the biggest contribution to infidelity).

\section{Numerical CNOT gate duration and single-qubit rotations}
\label{sec-num-CNOT}

 \begin{figure}[t]
\includegraphics[width=0.95\linewidth]{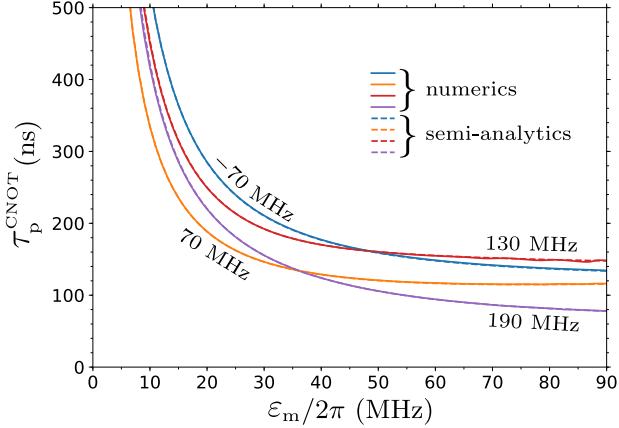}
 	\caption{CNOT gate duration $\tau_{\rm p}^{\rm \scriptscriptstyle CNOT}$ (neglecting single-qubit rotations) as a function of the mid-pulse drive amplitude $\varepsilon_\text{m}$ for several detunings: $\Delta/2\pi= -70$, 70, 130, and 190 MHz, while $g/2\pi = 3$ MHz, $\eta_{\rm c}/2\pi=\eta_{\rm t}/2\pi=300$ MHz, $\tau_{\rm r}/ \tau_{\rm p}=0.3$, and $\omega_{\rm d}=\omega_{\rm t}^{\rm c0}$. Solid lines are calculated numerically, dashed lines (almost coinciding with the solid lines) are calculated using the semi-analytical method.
 }
 	\label{fig-CNOT-time-num}
 \end{figure}

In numerical simulations, we use the qubit-qubit coupling $g/2\pi=3$ MHz (except in Fig.\ \ref{fig-g-dependence}) and transmon anharmonicity $\eta_{\rm c}/2\pi=\eta_{\rm t}/2\pi=300$ MHz. Figure \ref{fig-CNOT-time-num} shows the CNOT gate duration $\tau_{\rm p}^{\rm \scriptscriptstyle CNOT}$ as a function of the drive amplitude $\varepsilon_{\rm m}$ in the flat  middle part of a pulse (with $\tau_{\rm r}=0.3\,\tau_{\rm p}$), for several values of the qubit-qubit detuning: $\Delta/2\pi=-70$, 70, 130, and 190 MHz. For numerical results (solid lines) we choose $\omega_{\rm d}=\omega_{\rm t}^{\rm c0}$, i.e., $\delta =\omega_{\rm t}-\omega_{\rm t}^{\rm c0}$.
The dashed lines (almost coinciding with the solid lines) show the result of the semi-analytical method, in which we use Eq.\ (\ref{CNOT-basic}) and integrate over the pulse shape. We see that the semi-analytical method works very well; however, there are (barely) visible deviations at both small and large amplitudes $\varepsilon_{\rm m}$. We guess the slight deviation at large $\varepsilon_{\rm m}$ is because for a short pulse the non-adiabatic evolution during the ramps starts to play a noticeable role. The deviation at small $\varepsilon_{\rm m}$ is because here the $zz$-coupling (\ref{zz-coupling}) starts to play a relatively significant role.

\begin{figure}[t]
\includegraphics[width=0.95\linewidth]{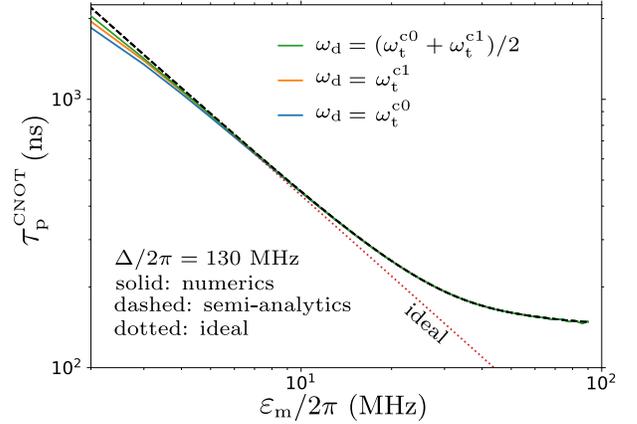}
	\caption{Solid lines: dependence of the CNOT duration $\tau_{\rm p}^{\rm \scriptscriptstyle CNOT}$ on $\varepsilon_{\rm m}$ for three drive frequencies:  $\omega_{\rm d}=\omega_{\rm t}^{\rm c0}$
(exact resonance for the control-qubit state $|0\rangle$, blue line), $\omega_{\rm d}=\omega_{\rm t}^{\rm c1}$ (resonance for the control-qubit state $|1\rangle$, orange line), and  $\omega_{\rm d}=(\omega_{\rm t}^{\rm c0}+\omega_{\rm t}^{\rm c1})/2$
(exactly in between, green line). The dashed line shows the semi-analytical results, straight dotted line corresponds to Eq.\ (\ref{CNOT-ideal-shape}).   We use $\Delta/2\pi=130$ MHz, other parameters are as in Fig.\ \ref{fig-CNOT-time-num}.
 }
	\label{fig-CNOT-time-num-2}
\end{figure}

For a more detailed analysis of the deviation at small $\varepsilon_{\rm m}$, solid lines in Fig.\ \ref{fig-CNOT-time-num-2} show the numerical CNOT time $\tau_{\rm p}^{\rm \scriptscriptstyle CNOT}$ for $\Delta/2\pi= 130$ MHz and three values of the drive frequency: $\omega_{\rm d}=\omega_{\rm t}^{\rm c0}$  (blue line, drive on-resonance with the target qubit when the control qubit is $|0\rangle$, i.e., $\delta =\omega_{\rm t}-\omega_{\rm t}^{\rm c0}$), $\omega_{\rm d}=\omega_{\rm t}^{\rm c1}$ (orange line, on-resonance when the control qubit is $|1\rangle$, i.e., $\delta =\omega_{\rm t}-\omega_{\rm t}^{\rm c1}$), and exactly in between, $\omega_{\rm d}=(\omega_{\rm t}^{\rm c0}+\omega_{\rm t}^{\rm c1})/2$ (green line).
The dashed line shows the semi-analytical result (actually, there are three dashed lines for the three values of $\delta$, but they are indistinguishable), and the dotted line shows the ideal result,
    \be
\tau_{\rm p,\, ideal}^{\rm \scriptscriptstyle CNOT}= \frac{\pi/2}{0.7\, \varepsilon_{\rm m}} \,\frac{\Delta (\eta_{\rm c}-\Delta)}{2g\eta_{\rm c} } ,
    \label{CNOT-ideal-shape}\ee
which follows from Eqs.\ (\ref{CNOT-basic}) and (\ref{tilde-varepsilon-1-0}) after integration over the pulse shape (\ref{pulse-shape}) with $\tau_{\rm r}/\tau_{\rm p}=0.3$ (this integration gives the factor 0.7). We see that the solid lines noticeably deviate down from the dashed line for $\varepsilon_{\rm m}/2\pi$ less than $\sim$5 MHz, with the largest deviation for $\omega_{\rm d}=\omega_{\rm t}^{\rm c0}$ (blue line). Note that for $\varepsilon /2\pi=5$ MHz, Eq.\ (\ref{tilde-varepsilon-1-0}) gives $(\tilde\varepsilon_1 -\tilde\varepsilon_0)/2\pi=0.41$ MHz, while $\omega_{zz}/2\pi= 0.15$ MHz [see Eq.\ (\ref{zz-coupling-an})], so the effect of $zz$-coupling is expected to be significant.

In more detail, for $\omega_{\rm d}=\omega_{\rm t}^{\rm c0}$  (blue line in Fig.\ \ref{fig-CNOT-time-num-2}) the drive is exactly on-resonance with the target qubit for the control-qubit state $|0\rangle$ and therefore the approximation $\varphi_0= 0.7 \, \tau_{\rm p}^{\rm \scriptscriptstyle CNOT} \times 2 \tilde\varepsilon_0 (\varepsilon_{\rm m})$ using Eq.\ (\ref{tilde-varepsilon-0}) should work well (as confirmed by numerics). In contrast, for the control-qubit state $|1\rangle$, the drive is detuned by $\omega_{zz}$ from the target-qubit frequency, which leads to Rabi oscillations with frequency $\sqrt{(2\tilde\varepsilon_1)^2+\omega_{zz}^2}$ within the plane tilted by angle ${\rm atan} (\omega_{zz}/2\tilde\varepsilon_1)$ from the $zy$ plane (the frequency and the plane are changing in time because of the pulse shape). The larger Rabi frequency (due to $\omega_{zz}$ contribution) leads to a slightly shorter CNOT gate duration than expected analytically, explaining the behavior of solid lines in Fig.\ \ref{fig-CNOT-time-num-2} at small $\varepsilon_{\rm m}$. The same effect (rotation within a tilted plane) leads to a significant infidelity of the CNOT gate at small $\varepsilon_{\rm m}$ -- see Sec.\ \ref{sec-err}.

\vspace{0.2cm}

As discussed above, in order to realize the CNOT operation, the CR gate with the pulse duration $\tau_{\rm p}^{\rm \scriptscriptstyle CNOT}$ should be complemented by single-qubit rotations. The target qubit should be rotated about $x$-axis by the angle $-\varphi_0$ to compensate the operator $e^{-i(\varphi_0/2)X_{\rm t}}$ in Eq.\ (\ref{U-class}). Similarly, the control qubit should be rotated about $z$-axis to compensate the relative phase $\theta_1-\theta_0$ and the negative imaginary unit due to the relation $e^{-i(\pi/2)X_{\rm t}}=-iX_{\rm t}$. So, naively we would expect that  $z$-rotation by the angle $\pi/2-(\theta_1-\theta_0)$ is needed. However, the angle $\theta_1-\theta_0$ is numerically computed in the rotating frame of the drive, while experimental $z$-rotation should be in the rotating frame of the control qubit. For the latter frame based on frequency $\omega_{\rm c}^{\rm t0}=E_{\overline{|1,0\rangle}}^{\rm (lf)}-E_{\overline{|0,0\rangle}}^{\rm (lf)}$ (i.e., when the target qubit is $|0\rangle$, as usually done in experiments), we need to replace the phase difference $\theta_1-\theta_0$ with
    \be
\theta_1'-\theta_0' = \theta_1-\theta_0 + (\omega_{\rm c}^{\rm t0}-\omega_{\rm d}) \,\tau_{\rm p}^{\rm \scriptscriptstyle CNOT},
    \ee
so in an experiment, the control qubit should be $z$-rotated by the angle $\theta_0'-\theta_1'+\pi/2$. Note that   $ \omega_{\rm c}^{\rm t0}-\omega_{\rm d}= \Delta + \delta +\omega_{\rm c}^{\rm t0}-\omega_{\rm c}$.

\begin{figure}[t]
\includegraphics[width=0.95\linewidth]{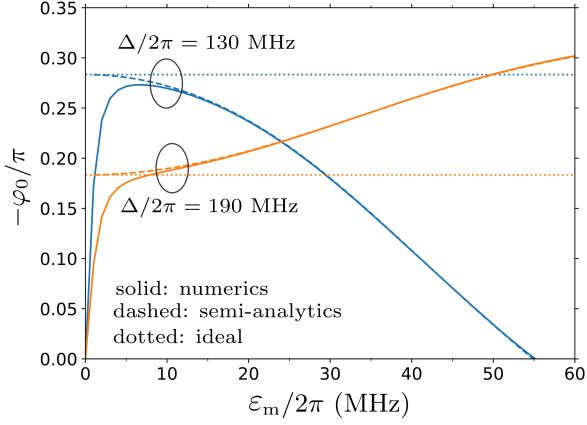}
	\caption{The angle $-\varphi_0$ of the compensating $x$-rotation of the target qubit to produce CNOT, as a function of the mid-pulse drive amplitude $\varepsilon_{\rm m}$. Numerical results are shown by solid lines, semi-analytics is represented by dashed lines, and dotted lines show the ideal result: $-\varphi_{0}/\pi=(\eta_{\rm c}-\Delta)/2\eta_{\rm c}$. Here we use $\Delta/2\pi=130$ and 190 MHz, $\omega_{\rm d}=\omega_{\rm t}^{\rm c0}$, $g/2\pi = 3$ MHz, $\eta_{\rm c}/2\pi=\eta_{\rm t}/2\pi=300$ MHz, and $\tau_{\rm r}/\tau_{\rm p}=0.3$.
 }
	\label{fig-Xrot}
\end{figure}

Figure \ref{fig-Xrot} shows the angle $-\varphi_0$ (normalized by $\pi$) as a function of $\varepsilon_{\rm m}$ for two values of the detuning: $\Delta/2\pi=130$ and 190 MHz (we use $\omega_{\rm d}=\omega_{\rm t}^{\rm c0}$). Solid lines show the numerical results, dashed lines show the corresponding semi-analytical results (integrating $- 2\tilde\varepsilon_0$ over the pulse shape), and horizontal dotted lines show the ideal result based on Eqs. (\ref{tilde-varepsilon-0}) and (\ref{tilde-varepsilon-1}): $-\varphi_0=\pi(\eta_{\rm c}-\Delta)/2\eta_{\rm c}$. We see that this ideal result for $-\varphi_0$ is never applicable. The deviation from it at large $\varepsilon_{\rm m}$ is described well by the semi-analytics and is due to the deviations from the ideal straight lines in Fig.\ \ref{fig-semi-analytics-comp}. The deviation from the ideal result in Fig.\ \ref{fig-Xrot} at small $\varepsilon_{\rm m}$ is due to $\omega_{zz}$ -- the same effect as discussed above for Fig.\ \ref{fig-CNOT-time-num-2}.

\begin{figure}[t]
\includegraphics[width=0.95\linewidth]{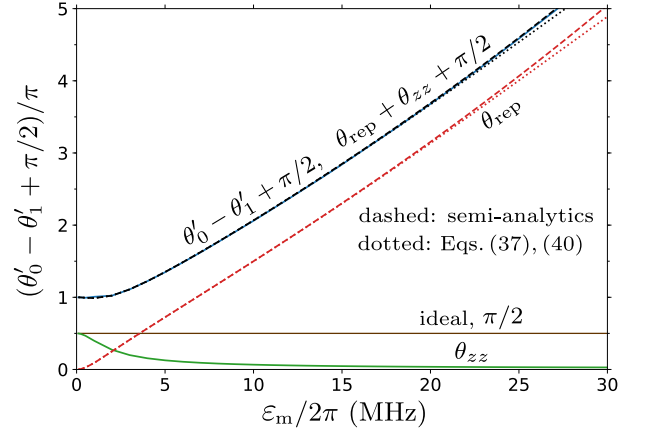}
	\caption{Upper (blue) solid line: numerical result for the angle $\theta_0'-\theta_1'+\pi/2$ of $z$-rotation of the control qubit, needed to produce the CNOT gate. The horizontal (brown) line shows the ideal value $\pi/2$. Dashed and dotted red lines show the contribution $\theta_{\rm rep}$ due to $\varepsilon$-induced level repulsion (\ref{theta-rep}), calculated either semi-analytically (dashed line) or via Eqs.\ (\ref{eigenenergy-0}) and (\ref{eigenenergy-1}) (dotted line). The lower (green) solid line shows the contribution $\theta_{zz}$ given by Eq.\ (\ref{theta-zz}). The upper solid and dashed black lines (very close to the blue solid line) show the sum $\theta_{\rm rep}+\theta_{zz}+\pi/2$. We use  $\Delta/2\pi=130$ MHz and parameters from Fig.\ \ref{fig-Xrot}.
  }
	\label{fig-Zrot}
\end{figure}

Upper (blue) solid line  in Fig.\ \ref{fig-Zrot} shows the numerical result for the control-qubit rotation angle $\theta_0'-\theta_1'+\pi/2$ (normalized by $\pi$) as a function of $\varepsilon_{\rm m}$ for $\Delta /2\pi= 130$ MHz and $\omega_{\rm d}=\omega_{\rm t}^{\rm c0}$. We see that it is quite different from the ideally expected value of $\pi/2$ (horizontal brown line). The difference is mainly  due to two effects. First, strong drive $\varepsilon (t)$ causes the level repulsion (ac Stark shift) in the control qubit, which slightly changes the control-qubit frequency and produces the accumulated angle
    \be
     \theta_{\rm rep} = \int_0^{\rm \tau_p} (E_{\overline{|1\rangle}_\varepsilon}-E_{\overline{|0\rangle}_\varepsilon}) \, dt .
    \label{theta-rep}\ee
(Actually, because the state $|1\rangle$ is assumed to be at the bottom of the Bloch sphere, this produces the $z$-rotation of $-\theta_{\rm rep}$, so we need to apply the rotation of $+\theta_{\rm rep}$ to compensate it.)   The angle $\theta_{\rm rep}$ calculated semi-analytically [see Eq.\ (\ref{Schrod-eq-control})] is shown by the dashed red line in Fig.\ \ref{fig-Zrot}; the same angle calculated using analytical results (\ref{eigenenergy-0}) and (\ref{eigenenergy-1}) is shown by the dotted red line (in both cases we use numerical $\tau_{\rm p}^{\rm \scriptscriptstyle CNOT}$ to integrate over the pulse shape).
The second contribution into $\theta_0'-\theta_1'$ is due to $zz$ coupling (\ref{zz-coupling}), which produces
    \be
\theta_{zz}= \frac{\omega_{zz}}{2}\, \tau_{\rm p}.
    \label{theta-zz}\ee
This is because the rotating frame here is based on $\omega_{\rm c}^{\rm t0}$, while $\omega_{\rm c}^{\rm t1}-\omega_{\rm c}^{\rm t0}=\omega_{zz}$ and both states of the target qubit participate equally (leading to the factor of $1/2$). The angle $\theta_{zz}$ is shown by the lower (green) solid line in Fig.\ \ref{fig-Zrot}. Adding the three contributions, $\theta_{\rm rep}+\theta_{zz}+\pi/2$, we obtain the dashed and dotted black lines (corresponding to the dashed and dotted red lines for $\theta_{\rm rep}$), which are quite close to the numerical result (solid blue line). This confirms the main physical mechanisms contributing to $\theta_0'-\theta_1'$ and also shows that the approximation based on Eqs.\  (\ref{eigenenergy-0}) and (\ref{eigenenergy-1}) works quite well.

\section{Error budget}
\label{sec-err}

In the previous section, we have discussed numerical results for the parameters of a CR-based CNOT gate: the duration $\tau_{\rm p}^{\rm \scriptscriptstyle CNOT}$ as a function of the mid-pulse drive amplitude $\varepsilon_{\rm m}$ and the compensating single-qubit rotation angles $-\varphi_0$ and $\theta_0'-\theta_1'+\pi/2$. We have seen that these parameters can be obtained quite accurately by the semi-analytical method. In this section we discuss numerical results for the infidelity $1-F_{MU}$ of the CR-based CNOT gate (also as a function of $\varepsilon_{\rm m}$), neglecting decoherence and infidelity of single-qubit rotations (see Sec.\ \ref{sec-numerical} for the definition of $F_{MU}$ and the calculation method). These results cannot be obtained semi-analytically and necessarily require full numerical simulations.

\begin{figure}[t]
\includegraphics[width=0.95\linewidth]{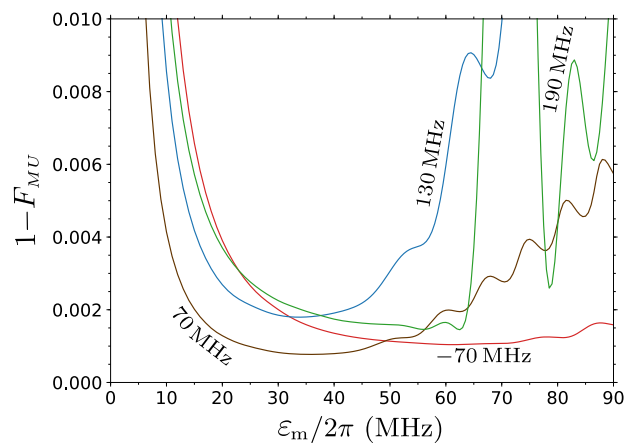}
	\caption{Infidelity $1-F_{MU}$ of the CNOT-equivalent gate as a function of mid-pulse drive amplitude $\varepsilon_\text{m}$ for several values of detuning: $\Delta/2\pi= -70$ MHz (red line), 70 MHz (brown line), 130 MHz (blue line), and 190 MHz (green line). Other parameters are $g/2\pi=3$ MHz, $\eta_{\rm c}/2\pi=\eta_{\rm t}/2\pi=300$ MHz, $\omega_{\rm d}=\omega_{\rm t}^{\rm c0}$, and $\tau_{\rm r}/\tau_{\rm p}=0.3$.
 }
	\label{fig-infidelity_plot}
\end{figure}

Figure \ref{fig-infidelity_plot} shows the numerical results for the CNOT gate infidelity $1-F_{MU}$ as a function of the mid-pulse drive amplitude $\varepsilon_{\rm m}$ for several values of the detuning: $\Delta/2\pi= -70$, 70, 130, and 190 MHz. As in the previous plots, we use $g/2\pi=3$ MHz, $\eta_{\rm c}=\eta_{\rm t}=300$ MHz, $\omega_{\rm d}=\omega_{\rm t}^{\rm c0}$, and $\tau_{\rm r}=0.3\, \tau_{\rm p}$. Most importantly, we see that the infidelity dependence on $\varepsilon_{\rm m}$ has a minimum, and at this minimum, the infidelity is crudely $10^{-3}$ for all lines. The second observation is that the minimum is not sharp and is reached for the values of the drive amplitude $\varepsilon_{\rm m}$ above which the CNOT duration $\tau_{\rm p}^{\rm \scriptscriptstyle CNOT}$ does not become significantly shorter by a further increase of  $\varepsilon_{\rm m}$ (see Figs.\ \ref{fig-speed-semi-analyt} and \ref{fig-CNOT-time-num}). Note that we do not take into account the effect of decoherence, which can be crudely (ideally) estimated by Eq.\ (\ref{Delta-F}). If decoherence were added, then the minima in Fig.\ \ref{fig-infidelity_plot} would shift to higher values of $\varepsilon_{\rm m}$; however, since the corresponding decrease of $\tau_{\rm p}^{\rm \scriptscriptstyle CNOT}$ is not significant, the benefit for fidelity is also not significant; moreover, using higher drive amplitudes could lead to other experimental problems. Therefore, we think the optimum values of $\varepsilon_{\rm m}$ in Fig.\ \ref{fig-infidelity_plot} should be somewhat close to experimental optima.

To understand the reason for the minima in Fig.\ \ref{fig-infidelity_plot} and to understand the physical mechanisms contributing to the infidelity (error budget), we have done the following more detailed calculations. We remind that we calculate fidelity $F_{MU}$ between the non-unitary $4\times 4$ matrix $M$ (which is the projection of the $35\times 35$ matrix of actual evolution onto the computational subspace) and the closest unitary matrix $U$, which belongs to the class (\ref{U-class}) with $\varphi_1-\varphi_0=\pi$ ($\rm mod\, 2\pi$) for a CNOT-equivalent gate. Now let us consider a bigger class of unitary matrices and define $\tilde M$ as the matrix, which is closest to $M$ out of
unitaries satisfying the condition
    \begin{align}
\tilde{M} = |0\rangle\langle 0|_\text{c}\, \tilde U_0^{\text{t}}+|1\rangle\langle 1|_\text{c}\, \tilde U_1^{\text{t}},
    \label{tilde-M-def}\end{align}
where $\tilde U_0^{\text{t}}$ and $\tilde U_1^{\text{t}}$ are any $2\times 2$ unitary matrices acting on the target qubit (here ``closest'' means that $\tilde M$ maximizes the fidelity $F_{M\tilde M}$). From this definition, $\tilde M$ does not change the control-qubit states $|0\rangle$ and $|1\rangle$ (i.e., does not allow any leakage from them), but its rotation of the target qubit is arbitrary. The matrix $U$ in Eq.\ (\ref{U-class}) is more restrictive in the sense that it allows only $x$-rotations of the target qubit.

\begin{figure}[t]
\includegraphics[width=0.95\linewidth]{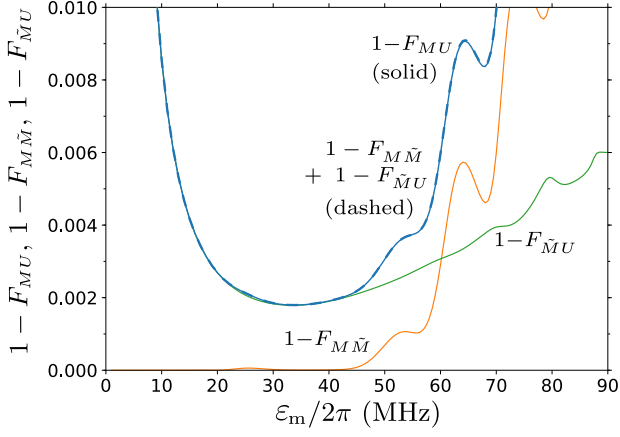}
	\caption{Decomposition of the CNOT gate infidelity $1-F_{MU}$ (blue solid line) into the leakage contribution $1-F_{M\tilde M}$ (orange line) and contribution $1-F_{\tilde M U}$ due to imperfect unitaries (green line). The sum $(1-F_{M\tilde M})+(1-F_{\tilde M U})$ (dashed blue line) is practically indistinguishable from the solid blue line. We use $\Delta/2\pi=130$ MHz and  parameters from Fig.\ \ref{fig-infidelity_plot}.
 }
	\label{fig-error_budget_plot}
\end{figure}

To clarify the error budget, we calculate additional infidelities $1-F_{M\tilde M}$ (between $M$ and $\tilde M$) and $1-F_{\tilde M U}$ (between $\tilde M$ and $U$) for the CNOT-equivalent CR operation. The idea is that the infidelity $1-F_{M\tilde M}$ is due to leakages (from the control-qubit states $|0\rangle$ and $|1\rangle$ to any state and also outside of the computational subspace for the target qubit). In contrast, the infidelity $1-F_{\tilde M U}$ is due to imperfect unitary rotations of the target qubit. From the physical approach of separation of the error into these different mechanisms, we would expect
    \begin{align}
1-F_{MU}\approx (1-F_{M\tilde{M}})+(1-F_{\tilde{M}U}).
  \label{decomposition}\end{align}
This is not an exact relation mathematically (the exact relation would require that certain elements of the quantum process tomography matrix for $M$ have exactly zero real and/or imaginary parts). However, numerical results (e.g., Fig.\ \ref{fig-error_budget_plot}) show that this relation is very accurate.

In Fig.\ \ref{fig-error_budget_plot}, the solid blue line, $1-F_{MU}$, is the same as the blue line in  Fig.\ \ref{fig-infidelity_plot} ($\Delta/2\pi = 130$ MHz). For the same case, the orange line shows $1-F_{M\tilde M}$, the green line shows $1-F_{\tilde M U}$, and the dashed blue line (practically indistinguishable from the solid blue line) shows the sum $(1-F_{M\tilde M})+(1-F_{\tilde M U})$. We see that the total infidelity $1-F_{MU}$ can really be decomposed into the infidelity $1-F_{M\tilde M}$ due to leakages and infidelity $1-F_{\tilde M U}$ due to imperfect unitaries acting on the target qubit. Similar decomposition into leakages and imperfect unitaries also works  well for other lines in Fig.\ \ref{fig-infidelity_plot} (relative inaccuracy of the decomposition is typically about $10^{-3}$).

Besides introducing the error budget via the decomposition (\ref{decomposition}), we also tried a further decomposition by introducing another $4\times 4 $ matrix $\tilde M'$, which is a  two-qubit unitary closest to $M$. In this case the infidelity $1-F_{M\tilde M'}$ is due to leakages outside of the computational subspace, while the infidelity $1-F_{\tilde M'\tilde M}$ is due to unitary transitions between states $|0\rangle$ and $|1\rangle$ of the control qubit (which in our terminology are also leakages for the CR gate). Then the infidelity consists of 3 components,
    \begin{align}
1-F_{MU}\approx (1-F_{M\tilde{M'}})+(1-F_{\tilde M'\tilde M})+(1-F_{\tilde{M}U}).
  \label{decomposition-2}\end{align}
We have checked that this relation is quite accurate numerically. However, for the cases we checked, the matrix $\tilde M'$ was very close to either $M$ or $\tilde M$ (depending on the detuning $\Delta$, which determines the strongest leakage process). Therefore, usually only two terms in Eq.\ (\ref{decomposition-2}) are significant, and this is why we will continue using the simpler error decomposition (\ref{decomposition}) below.

As seen in Fig.\ \ref{fig-error_budget_plot}, at small $\varepsilon_{\rm m}$ the CNOT gate infidelity is dominated by the imperfection of the unitaries $\tilde U_0^{\rm t}$ and $\tilde U_1^{\rm t}$ (contribution $1-F_{\tilde M U}$), while at large $\varepsilon_{\rm m}$  the leakage contribution $1-F_{M\tilde M}$ dominates (the same result for other detunings $\Delta$). This is because at small $\varepsilon_{\rm m}$ the effect of $zz$-coupling is important (as discussed in Sec.\ \ref{sec-num-CNOT}), while at large $\varepsilon_{\rm m}$ the ramps of the pulse become short and high, making the process significantly non-adiabatic and causing leakages.

\begin{figure}[t]
\includegraphics[width=0.95\linewidth]{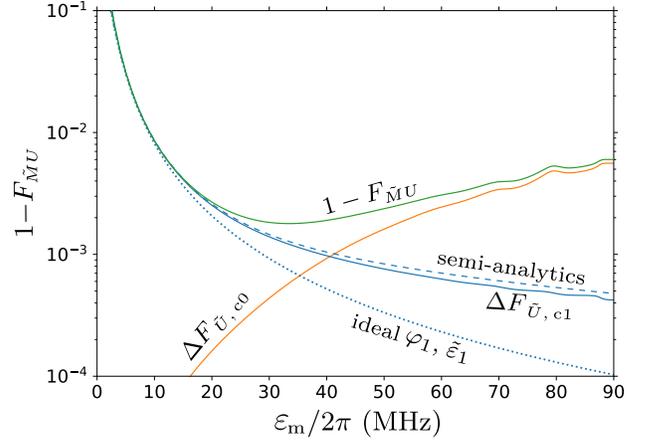}
	\caption{Further decomposition of the imperfect-unitary infidelity contribution $1-F_{\tilde M U}$ (green line) into contributions $\Delta F_{\tilde U,\, \rm c0}$ (orange line) and $\Delta F_{\tilde U,\, \rm c1}$ (blue solid line) for the control-qubit states $|0\rangle$ and $|1\rangle$, respectively. The dotted blue line shows analytical approximation for $\Delta F_{\tilde U,\, \rm c1}$ by Eq.\ (\ref{Delta-F-U1}) with ideal values for $\varphi_1$ and $\tilde\varepsilon_1$, while for the dashed blue line we use semi-analytical values for $\varphi_1$ and $\tilde\varepsilon_1$ in Eq.\ (\ref{Delta-F-U1}). Parameters are the same as in Fig.\ \ref{fig-error_budget_plot}.
 }
	\label{fig-improper_rotation}
\end{figure}

To clarify the dependence on $\varepsilon_{\rm m}$ of the imperfect-unitary contribution $1-F_{\tilde MU}$ (green line in Fig.\ \ref{fig-error_budget_plot}), we draw this line again in Fig.\ \ref{fig-improper_rotation} (now on semi-logarithmic scale). We  also show the numerical results for the contributions $\Delta F_{\,\tilde U, \, \rm c0}$ and $\Delta F_{\,\tilde U, \, \rm c1}$  to this line from imperfections of the unitaries $\tilde U_0^{\rm t}$ (for the control-qubit state $|0\rangle$) and $\tilde U_1^{\rm t}$ (for the control-qubit state $|1\rangle$): orange and blue solid lines in Fig.\ \ref{fig-improper_rotation}, respectively. [Mathematically, the definitions are: $\Delta F_{\,\tilde U, \, \rm c0}=4/5-(2/5) \, |{\rm Tr}(\tilde U_0^{\rm t} U_0^{\rm t\dagger})|$, $\Delta F_{\,\tilde U, \, \rm c1}= 4/5-(2/5) \, |{\rm Tr}(\tilde U_1^{\rm t} U_1^{\rm t\dagger})|$, where $U_0^{\rm t}$ and $U_1^{\rm t}$ are obtained from   Eq.\ (\ref{U-class}): $U_0^{\text{t}} = e^{-i(\varphi_0/2)X_{\rm t}}$ and $U_1^{\text{t}}=e^{-i(\varphi_1/2)X_{\rm t}}$.] At small  $\varepsilon_{\rm m}$ the main contribution comes from imperfect $\tilde U_1^{\rm t}$. This is because we use the drive frequency resonant with the target qubit for the control-qubit state $|0\rangle$ ($\omega_{\rm d}=\omega_{\rm t}^{\rm c0}$), so for the the control-qubit state $|1\rangle$ the drive is off resonance by $\omega_{zz}$ [see Eq. (\ref{zz-coupling})]. This detuning produces rotation of the target-qubit state about a tilted axis, instead of the desired $x$-axis. To check this explanation, we calculate analytically the corresponding infidelity,
    \be
    \Delta F_{\,\tilde U,\, \rm c1}  = \frac{2}{5} \, \sin^2 (\varphi_1/2) \, \frac{\omega_{zz}^2}{\frac{25}{40} [2\tilde\varepsilon_1 (\varepsilon_{\rm m})]^2+\omega_{zz}^2} ,
    \label{Delta-F-U1}\ee
where the factor $25/40$ comes from the integration of $\varepsilon^2(t)$ over the pulse shape (\ref{pulse-shape}) with $\tau_{\rm r}/\tau_{\rm p}=0.3$ (such integration in only the denominator is an approximation). The dotted blue line in Fig.\ \ref{fig-improper_rotation} shows this result with $\varphi_1$ and $\tilde\varepsilon_1$ calculated analytically: $\varphi_1 =(\eta_{\rm c}+\Delta)/2\eta_{\rm c}$ and Eq.\ (\ref{tilde-varepsilon-1}) for $\tilde\varepsilon_1$. The dashed blue line shows Eq.\ (\ref{Delta-F-U1}) with $\varphi_1$ and $\tilde\varepsilon_1$ calculated using the semi-analytical method. We see that both lines fit reasonably well the numerical result (solid blue line) at small $\varepsilon_{\rm m}$, with a better fit when using the semi-analytical values for $\varphi_1$ and $\tilde\varepsilon_1$. Even better fit (almost perfect, not shown) is when the Bloch-sphere evolution due to $\omega_{zz}$ and semi-analytical $\tilde\varepsilon (\varepsilon(t))$ is integrated over the pulse shape numerically, instead of using Eq.\ (\ref{Delta-F-U1}).

So, the contribution $\Delta F_{\,\tilde U,\, \rm c1}$ to the gate infidelity due to imperfect unitary $\tilde U_{\rm t}^{\rm c1}$ is well explained quantitatively. In contrast, we do not have a simple analytical way to find the contribution $\Delta F_{\,\tilde U,\, \rm c0}$ due to imperfect $\tilde U_{\rm t}^{\rm c0}$ (orange solid line in Fig.\ \ref{fig-improper_rotation}). Qualitatively, this contribution appears at large $\varepsilon_{\rm m}$ because the interplay between the level repulsions due to $g$ and $\varepsilon$  in Fig.\ \ref{fig3} slightly changes the frequency $\omega_{\rm t}^{\rm c0}$, so that there is no longer exact resonance with the drive, and the Bloch-sphere evolution also becomes tilted, thus producing the infidelity. Note that small oscillations of the orange and blue solid lines in Fig.\ \ref{fig-improper_rotation} at large $\varepsilon_{\rm m}$ are apparently related to the oscillations of the orange line in Fig.\ \ref{fig-error_budget_plot}, which are discussed below.

The second contribution to the overall gate infidelity $1-F_{MU}$ (Fig.\ \ref{fig-error_budget_plot}), which becomes dominating at large $\varepsilon_{\rm m}$, is the contribution $1-F_{M\tilde M}$
due to leakage produced by non-adiabaticity during the pulse ramps (orange line in Fig.\ \ref{fig-error_budget_plot}). From the fidelity definition (\ref{fidelity}), we expect an approximate relation
    \be
1-F_{M\tilde M} \approx P_{\rm leak}^{\rm out} + \frac{4}{5} P_{\rm leak}^{\rm comp},
    \label{leakage-2-comp}\ee
where $P_{\rm leak}^{\rm out}$ is the probability of leakage to outside the computational subspace, averaged over the initial two-qubit states, and $P_{\rm leak}^{\rm comp}$ is the averaged probability of leakage inside the computational subspace, which for our definition (\ref{tilde-M-def}) means transitions between states $|0\rangle$ and $|1\rangle$ of the control qubit. Note that to average the leakage probability over all initial two-qubit states, it is sufficient to average it over the 4 basis states: $\overline{|0,0\rangle}$, $\overline{|0,1\rangle}$, $\overline{|1,0\rangle}$, and $\overline{|1,1\rangle}$. We have checked the relation (\ref{leakage-2-comp}) numerically and it works very well; however, usually either the first or second term in Eq.\ (\ref{leakage-2-comp}) strongly dominates over the other term (depending on the dominating leakage channel).

\begin{figure}[t]
\includegraphics[width=0.95\linewidth]{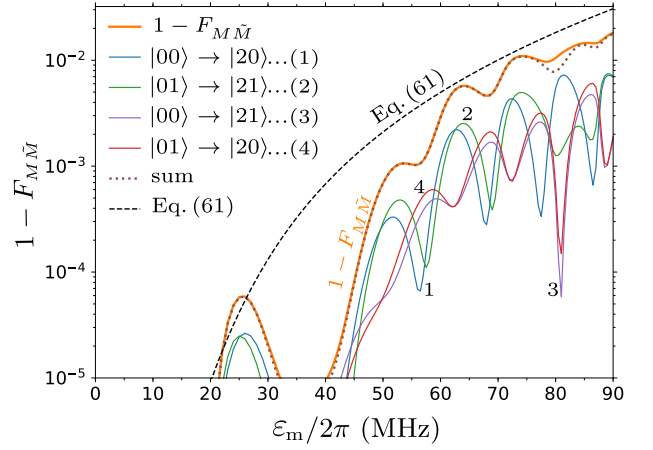}
	\caption{Numerical results for the leakage. The thick orange line shows infidelity contribution $1-F_{M\tilde M}$. Four thin solid lines (labeled 1, 2, 3, 4) show multiplied by the factor $1/4$ probabilities of the main leakage channels: $\overline{|0,0\rangle}\to \overline{|2,0\rangle}$, $\overline{|0,1\rangle}\to \overline{|2,1\rangle}$, $\overline{|0,0\rangle}\to \overline{|2,1\rangle}$, and $\overline{|0,1\rangle}\to \overline{|2,0\rangle}$. The sum of these four lines is shown by the brown dotted line; its closeness to the thick orange line verifies that the infidelity $1-F_{M\tilde M}$ is mainly due to these leakage channels. The black dashed line is the leakage probability estimate given by Eq.\ (\ref{leakage-ramp}). Parameters are the same as in Fig.\ \ref{fig-error_budget_plot}.
 }
	\label{fig-leakage}
\end{figure}

Figure \ref{fig-leakage} shows again the orange line $1-F_{M\tilde M}$ from Fig.\ \ref{fig-error_budget_plot} (now it is the thick orange line and the scale is semi-logarithmic) and also shows the numerical leakage probabilities (multiplied by the factor $1/4$ as in the averaging) for the processes $\overline{|0,0\rangle}\to \overline{|2,0\rangle}$, $\overline{|0,1\rangle}\to \overline{|2,1\rangle}$, $\overline{|0,0\rangle}\to \overline{|2,1\rangle}$, and $\overline{|0,1\rangle}\to \overline{|2,0\rangle}$ (thin solid lines). The sum of the thin solid lines is shown as the dotted brown line. It is very close to the thick orange line [as expected from Eq.\ (\ref{leakage-2-comp})], indicating that these are the four dominating leakage channels. A minor difference between the thick orange line and dotted brown line at $\varepsilon_{\rm m}$ close to 80 MHz is due to the significance of the additional leakage channels $\overline{|0,1\rangle}\to \overline{|1,2\rangle}$ and $\overline{|0,0\rangle}\to \overline{|1,2\rangle}$ (at this frequency the states $\overline{|2,1\rangle}$ and $\overline{|1,2\rangle}$  become on-resonance); similarly, a visible difference at about 40 MHz is because the transitions $|0\rangle \leftrightarrow |1\rangle$ in the control qubit become relatively important.

The main leakage channels in Fig.\ \ref{fig-leakage} involve the transition $|0\rangle \to |2\rangle$ in the control qubit. This is expected because in the rotating frame $E_0^{\rm (c)}-E_2^{\rm (c)}=\eta-2\Delta$ is only 40 MHz for the parameters of Fig.\ \ref{fig-leakage}. The matrix element of this transition via virtual state $|1\rangle$ is $-\sqrt{2}\,\varepsilon^2/\Delta$. Therefore, the non-adiabatic transition $\overline{|0\rangle_\varepsilon} \to \overline{|2\rangle_\varepsilon}$ during the front ramp with duration $\tau_{\rm r}$  of the pulse (\ref{pulse-shape}) has the amplitude proportional to the Fourier transform of $d(\varepsilon^2)/dt$ at the frequency $\eta-2\Delta$. The standard calculations give an estimate of the leakage probability in the control qubit during the front ramp:
    \be
      P_{\overline{|0\rangle_\varepsilon} \to \overline{|2\rangle_\varepsilon}} = \frac{2\pi^4\varepsilon_{\rm m}^4}{\Delta^2 (\eta-2\Delta)^6 \tau_{\rm r}^4}.
    \label{leakage-ramp}\ee
This probability is shown in Fig.\ \ref{fig-leakage} by the black dashed line. We see that it gives a reasonable (crude) approximation of the infidelity $1-F_{M\tilde M}$ due to leakage (if also multiplied by the factor of $1/4$, it goes close to the top of oscillating blue and green solid lines). Such a good fit is somewhat surprising because in deriving Eq.\ (\ref{leakage-ramp}) we assumed a fixed energy difference $E_0^{\rm (c)}-E_2^{\rm (c)}=\eta-2\Delta$, while for large $\varepsilon$ the difference of eigenenergies, $E_{\overline{|0\rangle_\varepsilon}} -E_{\overline{|2\rangle_\varepsilon}}$, becomes significantly larger, e.g., 60.7 MHz for $\varepsilon/2\pi=60$ MHz and 84.3 MHz for  $\varepsilon/2\pi=80$ MHz. Therefore, we would expect that Eq.\ (\ref{leakage-ramp}) should significantly overestimate the actual leakage (note the sixth power of $(\eta-2\Delta)$ in the denominator). As we checked, Eq.\ (\ref{leakage-ramp}) still works well because the non-adiabatic transition mainly accumulates during the lower half of the ramp, when $\varepsilon(t)$ is not too large.

The oscillations of the solid lines in Fig.\ \ref{fig-leakage} (leakage probabilities multiplied by $1/4$) are easily understandable. The non-adiabatic leakage occurs during both front and rear ramps, and the
transition amplitudes are added with a non-zero phase due to the energy difference $E_{\overline{|0\rangle_\varepsilon}} -E_{\overline{|2\rangle_\varepsilon}}$, accumulated between the ramps. So, the oscillations are due to constructive or destructive interference of the leakage contributions from the two ramps. We have checked that the $\varepsilon_{\rm m}$-difference between the peaks in Fig.\ \ref{fig-leakage} is consistent with estimates based on the numerical increase of $E_{\overline{|0\rangle_\varepsilon}} -E_{\overline{|2\rangle_\varepsilon}}$ with $\varepsilon$. The oscillations in the probabilities of individual leakage channels lead to the oscillations of their sum, thus explaining oscillations of $1-F_{M\tilde M}$ in Fig.\ \ref{fig-error_budget_plot} and oscillations of the overall CNOT gate infidelity $1-F_{MU}$ in Fig.\ \ref{fig-infidelity_plot} at large $\varepsilon_{\rm m}$.

Note that in Fig.\ \ref{fig-leakage} the leakage channels $\overline{|0,0\rangle}\to \overline{|2,0\rangle}$ and $\overline{|0,1\rangle}\to \overline{|2,1\rangle}$ (with non-changing state of the target qubit) have higher probabilities than for the leakage channels with changing state of the target qubit. This is because $|\varphi_0|\ll \pi$ in the interesting range of $\varepsilon_{\rm m}$ -- see Fig.\ \ref{fig-Xrot}, so the target-qubit state does not change much during the pulse when the control-qubit state is $|0\rangle$. In the opposite limit, $|\pi-\varphi_0|\ll \pi$, we would expect all four leakage channels to have approximately the same strength, and also would not expect significant oscillations (because the two ramps would mainly contribute to different channels).

Besides the detailed analysis of leakage contribution $1-F_{M\tilde M}$ for the detuning $\Delta/2\pi=130$ MHz, we have also analyzed the leakage for other values of the detuning in Fig.\ \ref{fig-infidelity_plot}. The case of  $\Delta/2\pi=190$ MHz is similar (the same dominating leakage channels) and Eq.\ (\ref{leakage-ramp}) still works well; however, $|\eta-2\Delta|$ is larger (80 MHz instead of 40 MHz), so the leakage becomes significant only at larger values of $\varepsilon_{\rm m}$.  Also, a very large value of the green line in  Fig.\ \ref{fig-infidelity_plot} at $\varepsilon_{\rm m}/2\pi \simeq 70$ MHz is due to an additional leakage channel $\overline{|0,1\rangle}\to \overline{|1,2\rangle}$, which is due to a resonance between these states. For the detuning $\Delta/2\pi$ of 70 MHz and $-70$ MHz (brown and red lines in Fig.\ \ref{fig-infidelity_plot}), the main leakage is between states $\overline{|0\rangle_\varepsilon}$ and $\overline{|1\rangle_\varepsilon}$ of the control qubit (also due to non-adiabaticity during the ramps).

Thus, in this section we have shown that in our model the error budget of the CR gate consists of two main contributions: the imperfection of the unitary operation at small drive amplitudes and the leakage at large drive amplitudes. At the optimal drive amplitude, the infidelity is on the order of $10^{-3}$. However, we did not take into account contributions from decoherence and also from possible problems caused by a strong drive of the control qubit (e.g., due to a resonance between an impurity and $E_{\overline{|0\rangle_\varepsilon}}\,- E_{\overline{|n\rangle_\varepsilon}}\,$  or $E_{\overline{|1\rangle_\varepsilon}}\,- E_{\overline{|n\rangle_\varepsilon}}$). Note that a strong leakage makes the CR gate operation impractical for the detuning $\Delta$ close to $0$, $\eta_{\rm c}/2$, $\eta_{\rm c}$, $3\eta_{\rm c}/2$, etc.

\section{Dependence on parameters}
\label{sec-parameters}


\begin{figure}[t]
\includegraphics[width=0.95\linewidth]{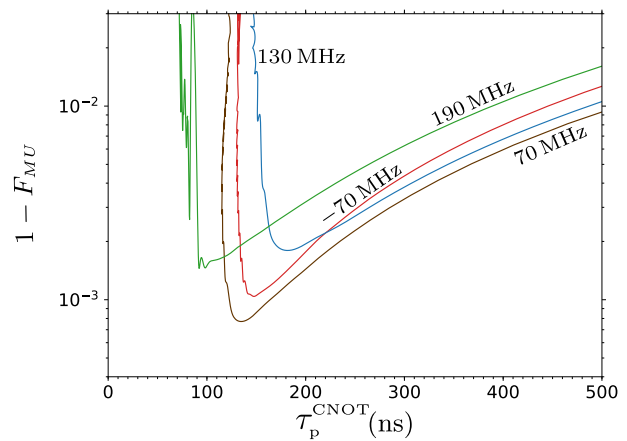}
	\caption{Parametric plot for the CNOT gate infidelity $1-F_{MU}$ versus the CNOT gate  duration $\tau_{\rm p}^{\rm \scriptscriptstyle CNOT}$ (the running parameter is $\varepsilon_{\rm m}$) for the detunings $\Delta/2\pi=-70$, 70, 130, and 190 MHz. We assume $g/2\pi=3$ MHz, $\eta_{\rm c}/2\pi=\eta_{\rm t}/2\pi=300$ MHz, $\omega_{\rm d}=\omega_{\rm t}^{\rm c0}$, and $\tau_{\rm r}/\tau_{\rm p}=0.3$.   }
	\label{fig-param-Delta}
\end{figure}

A convenient way to present numerical results \cite{Kirchhoff2018} is to parametrically plot infidelity $1-F_{MU}$ versus CNOT gate duration $\tau_{\rm p}^{\rm \scriptscriptstyle CNOT}$, with both quantities being functions of the drive amplitude $\varepsilon_{\rm m}$. Figure \ref{fig-param-Delta} presents such a plot for the results shown in Figs.\ \ref{fig-CNOT-time-num} and \ref{fig-infidelity_plot} for the detuning values $\Delta/2\pi=-70$, 70, 130, and 190 MHz, while other parameters are $g/2\pi=3$ MHz, $\eta_{\rm c}/2\pi=\eta_{\rm t}/2\pi=300$ MHz, $\omega_{\rm d}=\omega_{\rm t}^{\rm c0}$, and $\tau_{\rm r}/\tau_{\rm p}=0.3$. An increase of $\varepsilon_{\rm m}$ corresponds to moving from right to left along the lines. The main observation is that for each $\Delta$ the line is very steep at the left, which naturally corresponds to a limit on decreasing the duration of the CNOT gate. Among the plotted lines, the line for the detuning of 190 MHz gives the shortest duration (consistent with our discussion in Sec.\ \ref{sec-semi-analytics}); however, it does not give the best fidelity.


\begin{figure}[t]
\includegraphics[width=0.95\linewidth]{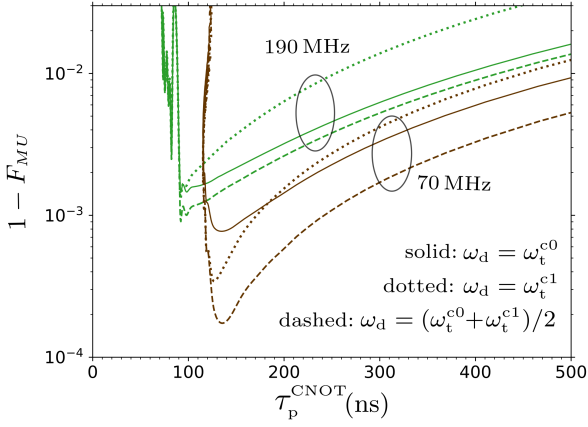}
	\caption{CNOT gate infidelity $1-F_{MU}$ versus duration  $\tau_{\rm p}^{\rm \scriptscriptstyle CNOT}$ (both are functions of $\varepsilon_{\rm m}$) for $\Delta/2\pi=70$ and 190 MHz and the drive frequency on-resonance with the target qubit for the control qubit either in the state $|0\rangle$ (solid lines, $\omega_{\rm d}=\omega_{\rm t}^{\rm c0}$) or in the state $|1\rangle$ (dotted lines, $\omega_{\rm d}=\omega_{\rm t}^{\rm c1}$) or exactly in between (dashed lines, $\omega_{\rm d}=(\omega_{\rm t}^{\rm c0}+\omega_{\rm t}^{\rm c1})/2$). Other parameters are as in Fig.\ \ref{fig-param-Delta}.
   }
	\label{fig-delta}
\end{figure}

In Fig.\ \ref{fig-delta} we show the results for three slightly different drive frequencies for the qubit detunings of 70 MHz (brown lines) and 190 MHz (green lines). Besides using the drive frequency resonant with the target qubit when the control-qubit state is $|0\rangle$ (solid lines, $\omega_{\rm d}=\omega_{\rm t}^{\rm c0}$, $\delta=\omega_{\rm t}-\omega_{\rm t}^{\rm c0}$), we also use the drive resonant with the target qubit when the control-qubit state is $|1\rangle$ (dotted lines, $\omega_{\rm d}=\omega_{\rm t}^{\rm c1}$), and also show the case of the drive frequency exactly in between (dashed lines, $\omega_{\rm d}=(\omega_{\rm t}^{\rm c0}+\omega_{\rm t}^{\rm c1})/2$). Note that $\omega_{\rm t}^{\rm c1}-\omega_{\rm t}^{\rm c0}=\omega_{zz}$ is 127 kHz and 200 kHz for the detunings of 70 MHz and 190 MHz, respectively [see Eq.\ (\ref{zz-coupling-an})].  We see that this small change of the drive frequency practically does not affect the natural limit for the duration $\tau_{\rm p}^{\rm \scriptscriptstyle CNOT}$; however, it may very significantly affect the infidelity. For example, for $\Delta/2\pi=70$ MHz the minimum infidelity is $1.7\times 10^{-4}$ for $\omega_{\rm d}=(\omega_{\rm t}^{\rm c0}+\omega_{\rm t}^{\rm c1})/2$, while it is $7.7\times 10^{-4}$  for $\omega_{\rm d}=\omega_{\rm t}^{\rm c0}$. We have checked that this small change of the drive frequency practically does not affect the leakage $1-F_{M\tilde M}$, but affects significantly the infidelity contribution $1-F_{\tilde M U}$ due to the imperfection of the unitary operation. This is exactly what is expected from the analysis in Sec.\ \ref{sec-err}, since the unitary imperfection is due to imperfect resonance between the drive and the target qubit. As we see from the numerical results, the drive frequency $\omega_{\rm d}=(\omega_{\rm t}^{\rm c0}+\omega_{\rm t}^{\rm c1})/2$ typically gives a better optimized fidelity than for $\omega_{\rm d}=\omega_{\rm t}^{\rm c0}$ or $\omega_{\rm d}=\omega_{\rm t}^{\rm c1}$.

\begin{figure}[b]
\includegraphics[width=0.95\linewidth]{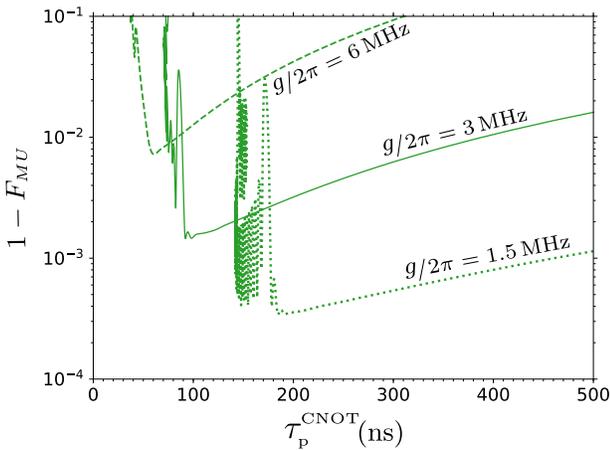}
	\caption{CNOT gate infidelity $1-F_{MU}$ versus duration  $\tau_{\rm p}^{\rm \scriptscriptstyle CNOT}$ for $\Delta/2\pi=190$ MHz and several values of the qubit-qubit coupling: $g/2\pi$=1.5, 3, and 6 MHz. Other parameters are as in Fig.\ \ref{fig-param-Delta}.
  }
	\label{fig-g-dependence}
\end{figure}

In all previous numerical plots, we assumed the qubit-qubit coupling $g/2\pi=3$ MHz. Figure \ref{fig-g-dependence} shows the CNOT gate infidelity for $g/2\pi=1.5$, 3, and 6 MHz, while $\Delta/2\pi=190$ MHz and other parameters are as in Fig.\ \ref{fig-param-Delta}. As expected, for minima of the lines in Fig.\ \ref{fig-g-dependence},  the CNOT gate duration decreases with increasing $g$ as $\tau_{\rm p}^{\rm \scriptscriptstyle CNOT}\propto g^{-1}$; however, we see that the infidelity increases crudely as $g^2$ (consistent with the scaling $\omega_{zz}\propto g^2$, so that $\omega_{zz}\tau_{\rm p}^{\rm \scriptscriptstyle CNOT}\propto g$). Increase of $g$ also increases undesired $zz$-interaction of idling qubits \cite{Galiautdinov2012} and therefore cannot be used as a simple way to reduce the CNOT gate duration in an experiment.

\begin{figure}[t]
\includegraphics[width=0.95\linewidth]{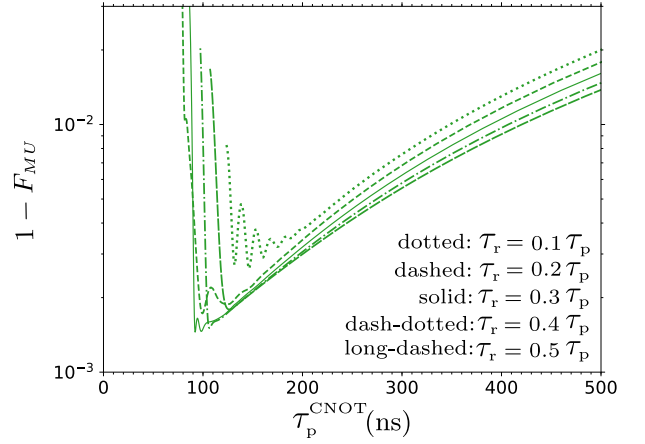}
	\caption{CNOT gate infidelity $1-F_{MU}$ versus duration  $\tau_{\rm p}^{\rm \scriptscriptstyle CNOT}$ for $\Delta/2\pi=190$ MHz and several values of the relative duration of the pulse ramps: $\tau_{\rm r}/\tau_{\rm p}=0.1$ (dotted line), 0.2 (dashed line), 0.3 (solid line), 0.4 (dash-dotted line), and 0.5 (long-dashed line).  Other parameters are as in Fig.\ \ref{fig-param-Delta}. The lines are cut at the left for clarity.
 }
	\label{fig-tau-ramp}
\end{figure}

In Fig.\ \ref{fig-tau-ramp} we numerically analyze dependence on the relative duration of the pulse ramp by changing the ratio $\tau_{\rm r}/\tau_{\rm p}$ in the pulse shape (\ref{pulse-shape}): $\tau_{\rm r}/\tau_{\rm p}=0.1$, 0.2, 0.3, 0.4, and 0.5 (other parameters are as in  Fig.\ \ref{fig-param-Delta} with $\Delta/2\pi=190$ MHz.) For clarity we do not show all oscillations on the left, cutting the lines at some maxima of the oscillations.  We see that at the right side of the graph (long $\tau_{\rm p}^{\rm \scriptscriptstyle CNOT}$) it is better to have the longest possible ramp, $\tau_{\rm r}/\tau_{\rm p}=0.5$. However, in the optimal range of short $\tau_{\rm p}^{\rm \scriptscriptstyle CNOT}$ with still small infidelity, the line with $\tau_{\rm r}/\tau_{\rm p}=0.3$ shows the best performance, which can be understood as the following trade-off. For longer ramps and the same $\tau_{\rm p}^{\rm \scriptscriptstyle CNOT}$, we need to use larger $\varepsilon_{\rm m}$ that increases the leakage (even though the ramp is smoother), while for shorter ramps and the same $\tau_{\rm p}^{\rm \scriptscriptstyle CNOT}$, the leakage is also increased because the ramp is too short and consequently non-adiabatic (even though $\varepsilon_{\rm m}$ is smaller). Thus, the ramps should be sufficiently smooth, but it is still beneficial to have a flat part of the pulse.

Finally, let us discuss the effect of the microwave crosstalk, which is always present in experiments \cite{Chow2011, Chow2012, Sheldon2016}. To include it, we need to add the crosstalk Hamiltonian
    \be
    H_{\rm ct} = \sum\nolimits_{n,m}  c_{\rm ct\,} \varepsilon (t) \sqrt{m}\, |n, m\rangle \langle n, m-1| + {\rm h.c.},
    \ee
which describes the microwave field applied directly to the target qubit. The crosstalk coefficient $c_{\rm ct}$ can in general be complex, and experimental results seem to indicate a complex $c_{\rm ct}$ \cite{Sheldon2016, Magesan2018}. However, for simplicity here we  assume a real $c_{\rm ct}$. In the ideal and semi-analytical theory, the crosstalk with real $c_{\rm ct}$ does not affect the results, except adding the phase $\varphi_{\rm ct} = \int_0^{\tau{\rm_p}} 2c_{\rm ct\,}\varepsilon (t)\, dt$ to both $\varphi_0$ and $\varphi_1$. However, it affects the numerical results because the crosstalk changes the Bloch-sphere angle of tilt caused by the effect of $\omega_{zz}$.

\begin{figure}[t]
\includegraphics[width=0.95\linewidth]{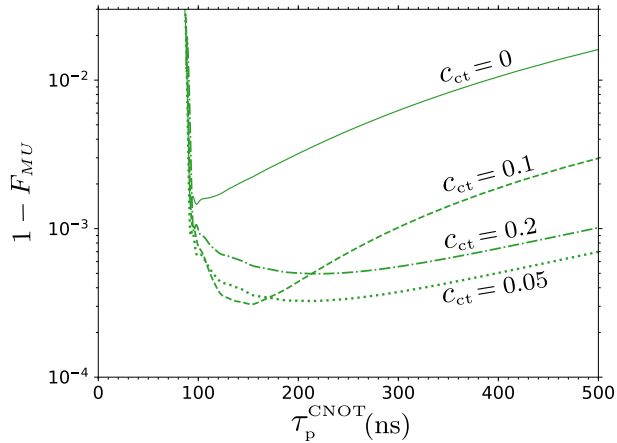}
	\caption{CNOT gate infidelity $1-F_{MU}$ versus duration  $\tau_{\rm p}^{\rm \scriptscriptstyle CNOT}$ for $\Delta/2\pi=190$ MHz and several values of the microwave crosstalk coefficient: $c_{\rm ct}=0$ (solid line), 0.05 (dotted line), 0.1 (dashed line) and 0.2 (dash-dotted line). Other parameters are as in Fig.\ \ref{fig-param-Delta}. The lines are cut at the left for clarity.
}
	\label{fig-crosstalk}
\end{figure}

Figure \ref{fig-crosstalk} shows numerical results for several values of the (real) crosstalk coefficient $c_{\rm ct}$ (the parameters are as in Fig.\ \ref{fig-param-Delta} with $\Delta/2\pi=190$ MHz). We see that the crosstalk improves fidelity. This improvement can be easily understood. The crosstalk does not affect leakage, but it decreases the imperfect-unitary contribution $1-F_{M\tilde M}$ because now the drive detuning $\omega_{\rm d}-\omega_{\rm t}^{\rm c1}$ should be compared with a larger Rabi frequency $\tilde\varepsilon_1+c_{\rm ct}\varepsilon$  instead of $\tilde\varepsilon_1$, and therefore the resulting tilt of the Bloch-sphere evolution is smaller (the same effect for the control-qubit state $|0\rangle$ if the signs of $\tilde\varepsilon_0$ and $c_{\rm ct}\varepsilon$ coincide). Thus, somewhat unexpectedly, the microwave crosstalk can improve the gate fidelity. However, this improvement is significant only if the effect of $\omega_{zz}$ is significant. For example, if in an experiment the infidelity is mainly determined by leakage and decoherence, then the crosstalk will not improve fidelity. Also, we remind that here we considered only real crosstalk coefficients, and the results for a complex $c_{\rm ct}$ may be significantly different.

The effect of the echo sequence is analyzed numerically in the Appendix. The main observation is that the echo-CR gate is typically   longer than the  basic CR gate for the same level of infidelity (even assuming instantaneous single-qubit gates).

\section{Conclusion}
\label{sec-conclusion}

In this paper, we have analyzed analytically, semi-analytically, and numerically the operation of the Cross-Resonance gate for superconducting qubits,  focusing on using the CR gate to realize the CNOT operation \cite{Chow2011,  Corcoles2015, Sheldon2016}. Our model has been based on the Hamiltonian (\ref{Ham})--(\ref{Ham-varepsilon}) and simple pulse shape (no echo sequence).

The analytical theory gives Eq.\ (\ref{tilde-varepsilon-n-res}) for the effective drive amplitude $\tilde\varepsilon_n$ of the target qubit, which depends on the control-qubit state $|n\rangle$. However, the analytics can be used only for sufficiently small (and not too small) physical drive amplitude $\varepsilon$. The next-order analytics (Sec.\ \ref{sec-next-order}) slightly widens the applicability range; however, it is still inapplicable  in the practically interesting range of $\varepsilon$.

We have found that the speed of the CR gate and the compensating single-qubit rotations can be obtained very accurately by a sufficiently simple semi-analytical theory, discussed in Sec.\ \ref{sec-semi-analytics}. This theory [Eqs.\ (\ref{Schrod-eq-control})--(\ref{tilde-epsilon-n-1q})] is based on solving a one-qubit time-independent Schr\"odinger equation. The semi-analytical theory depends on
only two parameters: dimensionless qubit-qubit detuning $\Delta/\eta_{\rm c}$ (normalization is the control-qubit anharmonicity $\eta_{\rm c}$) and dimensionless drive amplitude $\varepsilon/\eta_{\rm c}$. The dimensionless speed of the CR gate as a function of these two parameters is shown in Fig.\ \ref{fig-speed-semi-analyt}. The CR gate speed cannot be increased indefinitely by increasing the drive amplitude $\varepsilon$; the maximum speed (as a function of $\Delta/\eta_{\rm c}$) and the corresponding drive amplitude are shown in Fig.\ \ref{fig-speed-eps-max}. As follows from Figs.\ \ref{fig-speed-semi-analyt} and \ref{fig-speed-eps-max}, the best operation of the CR gate is expected for the detuning within the range $0.5\, \eta_{\rm c} <\Delta <  \eta_{\rm c}$.

Full numerical approach (discussed in Sec.\ \ref{sec-numerical}) is mainly needed to calculate intrinsic fidelity of the CNOT-equivalent CR gate.
The numerical results depend on the pulse shape, for which we use the simple cosine-ramp model, Eq.\ (\ref{pulse-shape}).
Most importantly, the infidelity $1- F_{MU}$ has a minimum as a function of the mid-pulse drive amplitude $\varepsilon_{\rm m}$. The minimum value depends on the detuning $\Delta$ (Figs.\ \ref{fig-infidelity_plot} and \ref{fig-param-Delta}), and for typical parameters used in this paper the optimal infidelity is on the order of $10^{-3}$ (though it approaches $10^{-4}$ for some parameters, and in principle, the theoretical infidelity can be arbitrarily small for complicated pulse shapes \cite{Kirchhoff2018}).

Our model does not include decoherence, so the error budget of the gate consists of two contributions: due to imperfect unitary operations and due to leakage [Eq.\ (\ref{decomposition})]. The imperfect unitary dominates at small $\varepsilon_{\rm m}$. The mechanism of this imperfection is related to the $zz$-interaction of the qubits [Eq.\ (\ref{zz-coupling})], which makes the target-qubit frequency dependent on the control-qubit state ($\omega_{\rm t}^{\rm c1}-\omega_{\rm t}^{\rm c0}=\omega_{zz}$), and therefore makes it impossible to use a drive frequency exactly on-resonance with the target qubit in both cases (for the control-qubit states $|0\rangle$ and $|1\rangle$). Because of this contribution into the error budget, at small $\varepsilon_{\rm m}$ the gate infidelity is very sensitive to small changes of the drive frequency (Fig.\ \ref{fig-delta}), and the microwave crosstalk can improve fidelity (Fig.\ \ref{fig-crosstalk}).

The other contribution to the error budget, which dominates at large drive amplitudes $\varepsilon_{\rm m}$, is due to leakages. The main leakage is in the strongly-driven control qubit and is caused by non-adiabaticity during the ramps of the pulse. Depending on the ratio $\Delta/\eta_{\rm c}$ between the detuning and control-qubit anharmonicity, the main leakage mechanism can be either $|0\rangle \leftrightarrow |1\rangle$ (between the control-qubit states $|0\rangle$ and $|1\rangle$) or $|0\rangle \to |2\rangle$ or some other leakage channel. In particular, for $|\Delta|\ll \eta_{\rm c}$ there is a strong leakage $|0\rangle \leftrightarrow |1\rangle$ because the drive frequency is near resonance with the control qubit. Similarly, for  $|\Delta-\eta_{\rm c}/2|\ll \eta_{\rm c}$ there is a strong leakage $|0\rangle \to |2\rangle$ because in this case  the energy difference between states $|2\rangle$ and $|0\rangle$ of the control qubit is near resonance with doubled frequency of the drive. In some cases we also observed a significant leakage for the channel $\overline{|0,1\rangle} \to \overline{|1,2\rangle}$ (when it becomes near-resonance with the doubled drive frequency). 

Strong leakage makes the CR gate operation impractical when detuning $\Delta$ is close to $0$, $\eta_{\rm c}/2$, $\eta_{\rm c}$, $3\eta_{\rm c}/2$, etc. Therefore, we would expect the best operation of the CR gate for the detuning within the range  $0.6\, \eta_{\rm c}<\Delta <0.8\,\eta_{\rm c}$  (see Figs.\ \ref{fig-speed-semi-analyt} and \ref{fig-speed-eps-max}). Another reasonable range (which requires smaller drive amplitudes) is $0.2\, \eta_{\rm c}<\Delta <0.3\,\eta_{\rm c}$.

Crudely, the CNOT-equivalent gate duration $\tau_{\rm p}^{\rm \scriptscriptstyle CNOT}$ (optimized over $\varepsilon_{\rm m}$) is comparable to $\pi/g$, with a coefficient  somewhat larger or smaller than 1, depending on $\Delta/\eta_{\rm c}$ -- see Fig.\ \ref{fig-param-Delta}. The optimized intrinsic infidelity for a simple pulse shape (\ref{pulse-shape}) is crudely comparable to $(g/\eta_{\rm c})^2$ (as follows from scaling of $\omega_{zz}$ and $\tau_{\rm p}^{\rm \scriptscriptstyle CNOT}$), with a coefficient typically about $10^0$--$10^{1.5}$, depending on the drive frequency, $\Delta/\eta_{\rm c}$, crosstalk,  etc. We have found that to reduce the non-adiabatic leakage, the ramps of the pulse should occupy a significant fraction of the pulse duration; however, the flat part in the middle of the pulse should not be shortened to zero.

In this paper we have not focused on analyzing the echo sequence. However, some numerical results for the echo-CR gate are presented in the Appendix. In the echo sequence, there are four ramps instead of two; therefore, for the same total pulse duration, the ramps are shorter. This increases non-adiabatic leakages, so for the same infidelity, the echo-CR gate duration is typically longer than for the basic CR gate (Fig.\ \ref{fig-cnot-delta-echo}).

A crude estimate of the CR gate infidelity contribution due to decoherence (within the computational subspace only) is given by Eq.\ (\ref{Delta-F}). However, for large drive amplitudes, the bare state $|2\rangle$ and higher states of the control qubit are significantly occupied; their decoherence is typically much faster than in the computational subspace and therefore can significantly increase the gate infidelity. Another potential mechanism for experimental CR gate infidelity is due to two-level systems (TLSs) produced by impurities. A strong drive changes the effective energy levels in the control qubit by about one hundred MHz, and therefore the TLSs can become on-resonance with the control qubit during the ramp of the pulse. Moreover, large drive amplitude makes multi-photon processes easily possible, and therefore TLSs can become resonant with combinational frequencies for many channels (similar to the situation for fast measurement of a qubit).

Note that in this paper we did not explicitly take into account the resonator, providing coupling between the qubits; instead we replaced it with an equivalent direct coupling. It would be interesting to repeat our numerical simulations, taking into account the resonator levels explicitly. However, we do not expect a significant modification of the results because the additional non-adiabatic effects should be suppressed by typically large detuning between the resonator and qubits. It would also be interesting to include decoherence into the simulations; however, it is not obvious what a proper model is for decoherence of higher levels, relevant to actual experimental situations.

We hope that some experimental group will carry out detailed measurements of the CNOT gate duration and error budget of the CR gate as functions of the drive amplitude, drive frequency, detuning, and pulse shape. It will be interesting (and important for the CR gate application in quantum computing) to compare experimental results with our theoretical findings.

\acknowledgments

The authors thank Juan Atalaya, Leonid Pryadko, Vinay Ramasesh, Machiel Blok, Ravi Naik, Irfan Siddiqi, Matthew Ware, and Britton Plourde for useful discussions. The work was supported by ARO grant No. W911NF-18-1-0178.

\appendix
\section{Effect of the echo sequence}
\label{sec-echo}

In the main text we focused on analysis of the basic CR gate. Here we analyze the echo-CR gate \cite{Sheldon2016, Corcoles2013, Takita2016}.

The echo-CR gate \cite{Corcoles2013} is essentially a sequence of two same-shape basic CR gates with halved rotation angles, $|\varphi_1-\varphi_0|=\pi/2$, with the control qubit state flipped ($|0\rangle \leftrightarrow |1\rangle$) in between the two halves of the procedure and with the flipped phase ($\varepsilon \to -\varepsilon $) of the applied microwave drive for the second half of the procedure. In this case the $x$-rotation angles for the target qubit become $\varphi_1=\varphi_1^{(1)}+\varphi_0^{(2)}=\varphi_1^{(1)}-\varphi_0^{(1)}$ and $\varphi_0=\varphi_0^{(1)}+\varphi_1^{(2)}=\varphi_0^{(1)}-\varphi_1^{(1)}$, where the superscripts refer to the first or second half of the procedure. Consequently, $\varphi_1=-\varphi_0$ and also $\varphi_1-\varphi_0=\pi$  when $\varphi_1^{(1)}-\varphi_0^{(1)}=\pi/2$. This is what is often called $ZX_{\pi/2}$ gate \cite{Chow2011, Corcoles2013, Sheldon2016}. Note that the relations $\varphi_0^{(2)}=-\varphi_0^{(1)}$ and $\varphi_1^{(2)}=-\varphi_1^{(1)}$ are because of the symmetry of the procedure and the phase shift by $\pi$ for the drive, $\varepsilon (t+\tau_{\rm p}/2) = -\varepsilon (t)$, where $\tau_{\rm p}/2$ is the time difference between the two halves of the procedure. Also note that the control qubit should be flipped back after the second pulse (though this flip can sometimes be compiled into the overall sequence of an algorithm).

Because of the symmetry, the echo sequence eliminates the need to apply the compensating $x$-rotation of the target qubit (by $-\varphi_0$, as assumed in the main text). It also eliminates the need to apply compensating $z$-rotation of the control qubit (by $\theta_0-\theta_1+\pi/2$, as assumed in the main text). This significantly reduces experimental complexity.   Nevertheless, if we want to produce CNOT gate from the echo-CR gate $ZX_{\pi/2}$, we still need to apply additional $x$-rotation by $\pi/2$ for the target qubit and $z$-rotation by $\pi/2$ for the control qubit.

The analytical (Sec.\ \ref{sec-ideal}) and semi-analytical theory (Sec.\ \ref{sec-semi-analytics}) for the echo-CR gate does not change compared with the basic CR gate because of the symmetry: we can just use the total gate duration with the same drive amplitude. In particular, Figs.\ \ref{fig-speed-semi-analyt} and \ref{fig-speed-eps-max} are still applicable without any change. However, numerical results are different because the ramps for the  echo-CR gate of the same duration are shorter and correspondingly the leakage is typically bigger.

In the numerical simulations, for the first half of the procedure ($0\leq t\leq \tau_{\rm p}/2$) we use the pulse shape (\ref{pulse-shape}) with substitutions $\tau_{\rm p}\to \tau_{\rm p}/2$ and $\tau_{\rm r}\to \tau_{\rm r}/2$, while for the second half ($\tau_{\rm p}/2 \leq t\leq \tau_{\rm p}$) we use the inverted shape, $\varepsilon(t)=-\varepsilon (t-\tau_{\rm p}/2)$. Therefore, $\tau_{\rm p}$ is the total pulse duration and $\tau_{\rm r}$ is the total duration of the two front ramps (or two rear ramps). We assume ideal instantaneous $\pi$-rotations of the control qubit (about $x$ axis) at time moments $\tau_{\rm p}/2$ and $\tau_{\rm p}$, and also ideal instantaneous rotations converting the $ZX_{\pi/2}$ gate into CNOT: $x$-rotation by $\pi/2$ for the target qubit and $z$-rotation by $\pi/2$ for the control qubit. The fidelity is calculated using Eq.\ (\ref{fidelity}), which compares the actual gate with the ideal $ZX_{\pi/2}$ gate, i.e., Eq.\ (\ref{U-class}) with $\theta_0-\theta_1=0$ and $|\varphi_1|=\pi/2$ (the relation $|\varphi_1-\varphi_0|=\pi$ is achieved by varying $\tau_{\rm p}$).

\begin{figure}[t]
\includegraphics[width=0.95\linewidth]{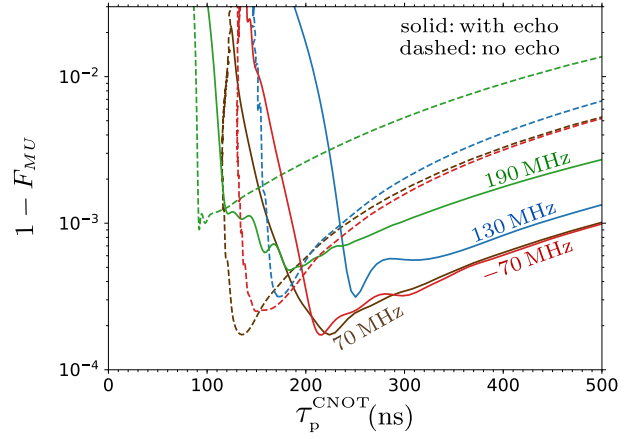}
	\caption{CNOT gate infidelity $1-F_{MU}$ versus duration $\tau_{\rm p}^{\rm \scriptscriptstyle CNOT}$ for the echo-CR gate (solid lines) and the basic CR gate (dashed lines). The detunings $\Delta/2\pi$ are $190$ MHz (green lines), 130 MHz (blue lines), 70 MHz (brown lines), and $-70$ MHz (magenta lines). The parameters are the same as in  Fig.\ \ref{fig-param-Delta}, except
we use $\omega_{\rm d}=(\omega_{\rm t}^{\rm c0}+\omega_{\rm t}^{\rm c1})/2$. The lines are cut at the left for clarity.
	}
	\label{fig-echo}
\end{figure}

Figure \ref{fig-echo} shows a comparison between the CR gate performances with and without echo sequence. The dashed lines (without echo) are the same as lines in Fig.\ \ref{fig-param-Delta}, except now we use  the drive frequency $\omega_{\rm d}=(\omega_{\rm t}^{\rm c0}+\omega_{\rm t}^{\rm c1})/2$. The solid lines show the results for the echo-CR gate with the same parameters (colors correspond to particular detunings $\Delta$).  Most importantly, we see that the steep increase at the left for the solid lines occurs at larger $\tau_{\rm p}^{\rm \scriptscriptstyle CNOT}$. This means that for the same infidelity $1-F_{MU}$, the echo-CR gate is longer than the basic CR gate (even not including the durations of the additional $\pi$-pulses). As mentioned above, this is because the leakage is a more severe problem for the echo-CR gate: four ramps instead of two make their durations shorter, and this significantly increases nonadiabaticity during ramps.

\begin{figure}[t]
\includegraphics[width=0.95\linewidth]{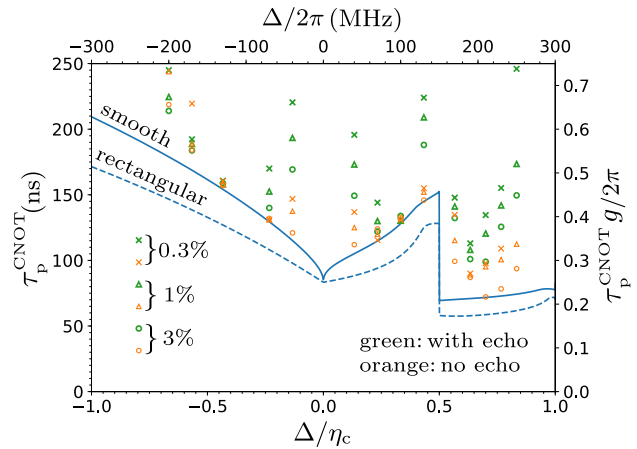}
	\caption{Symbols: numerical results for the CNOT gate duration $\tau_{\rm p}^{\rm \scriptscriptstyle CNOT}$ for the infidelity levels of $0.003$ (crosses), $0.01$ (triangles) and $0.03$ (circles), for several values of the detuning $\Delta$ (horizontal axis): $-200$, $-170$, $-130$, $-70$, $-40$, $40$, $70$, $100$, $130$, $170$, $190$, $210$, $230$, and $250$ MHz. Green (thicker) symbols are for the echo-CR gate, orange (thinner) symbols are for the basic CR gate. We use parameters $g/2\pi=3$ MHz, $\eta_{\rm c}/2\pi=\eta_{\rm t}/2\pi=300$ MHz, and $\tau_{\rm r}/\tau_{\rm p}=0.3$ (results on the dimensionless scales should not depend much on these parameters, except for  $\tau_{\rm r}/\tau_{\rm p}$). Solid line: results of the semi-analytical theory (optimized over the drive amplitude) for the same pulse shape. Dashed line: semi-analytical results for a rectangular pulse (the same as in Fig.\ \ref{fig-speed-eps-max}, but the vertical scale is inverted). Both dimensionless and dimensional scales are given for the axes.
}
	\label{fig-cnot-delta-echo}
\end{figure}

Figure \ref{fig-cnot-delta-echo} summarizes our semi-analytical and numerical results for the basic CR gate and the echo-CR gate. It shows the duration of the CNOT-equivalent gate $\tau_{\rm p}^{\rm \scriptscriptstyle CNOT}$ versus the detuning $\Delta$ between the control and target qubits, using both the dimensionless and dimensional scales. We use the coupling $g/2\pi= 3$ MHz, anharmonicity $\eta_{\rm c}=\eta_{\rm t}=300$ MHz, relative ramp duration $\tau_{\rm r}/\tau_{\rm p}=0.3$, and the drive frequency $\omega_{\rm d}=(\omega_{\rm t}^{\rm c0}+\omega_{\rm t}^{\rm c1})/2$. For different values of $g$, $\eta_{\rm c}$ and $\eta_{\rm t}$, the numerical results on the dimensionless scale are not expected to change significantly (semi-analytical results would remain exactly the same).

The dashed line in Fig.\  \ref{fig-cnot-delta-echo} shows the CNOT gate duration (minimized over the drive amplitude $\varepsilon$) as follows from the semi-analytical theory with a rectangular pulse shape. This line is the same as the solid line in Fig.\ \ref{fig-speed-eps-max} on the inverted  scale (duration instead of speed). The solid line in Fig.\  \ref{fig-cnot-delta-echo} shows the optimized CNOT gate duration for the smooth pulse shape (\ref{pulse-shape}) with $\tau_{\rm r}/\tau_{\rm p}=0.3$, also obtained semi-analytically. We see that this duration is longer than for the rectangular pulse, but not by the naive factor $1/(1-\tau_{\rm r}/\tau_{\rm p})=1.43$ (the ratio is significantly less than this factor because of the pulse-shape integration over the lines in Fig.\ \ref{fig-speed-semi-analyt}, for which the region near maximum is most important). The semi-analytical solid and dashed lines are the same for the  echo-CR and basic CR gates.

The symbols in  Fig.\  \ref{fig-cnot-delta-echo} show numerical durations of the CNOT-equivalent gates for three levels of the infidelity $1-F_{MU}$: 0.3\% (crosses), 1\% (triangles), and 3\% (circles), using the pulse shape (\ref{pulse-shape}) with $\tau_{\rm r}/\tau_{\rm p}=0.3$. Green (thicker) symbols are for the echo-CR gate, while orange (thinner) symbols are for the basic CR gate. The symbols are presented for the detunings $\Delta/2\pi$ of $-200$, $-170$, $-130$, $-70$, $-40$, $40$, $70$, $100$, $130$, $170$, $190$, $210$, $230$, and $250$ MHz. As expected, all symbols are above the  solid line. For the detunings of $-130$ and 100 MHz, the symbols are quite close to the solid line. This means that at these detunings, the leakage is not yet too strong for the near-optimal values of the drive amplitudes (e.g., compare lines in Figs.\ \ref{fig-CNOT-time-num} and \ref{fig-infidelity_plot} for $-70$ MHz);  the reason for a weak leakage is discussed later. In contrast, for some detunings, the symbols  in  Fig.\  \ref{fig-cnot-delta-echo} are much above the solid line. For example, for the detuning of 170 MHz, the semi-analytics predicts the duration of 70 ns, while for 1\% infidelity, the numerical results give 115 ns for the basic CR gate and 141 ns for the echo-CR gate. Such a big difference indicates a very significant leakage for the near-optimum drive amplitudes. As expected, in this case the echo-CR gate requires a significantly longer duration than the basic CR gate for the same level of infidelity.

Figure \ref{fig-cnot-delta-echo} can be used to estimate the range of best detunings, which provide close-to-shortest CNOT gate durations. For the basic CR gate with 1\% infidelity, the best detuning range is crudely $0.6 <\Delta /\eta_{\rm c} < 0.65$, which for our parameters provides the fastest CNOT gate duration of around 90 ns (excluding single-qubit pulses). For the echo-CR gate with 1\% infidelity, the best range is the same, and it provides the fastest duration of about 110 ns. Another reasonably well performing range of detunings is around $\Delta /\eta_{\rm c} \simeq 0.25$; for our parameters it gives CNOT gate duration of about 120 ns for the basic CR gate and 130 ns for the echo-CR gate (for 1\% infidelity). One more reasonable range is around $\Delta /\eta_{\rm c} \simeq - 0.25$; the corresponding CNOT gate durations are 130 ns (without echo) and 150 ns (with echo).

The CNOT gate duration is typically longer if we require a smaller infidelity. Correspondingly, the crosses in Fig.\ \ref{fig-cnot-delta-echo} typically are significantly higher than triangles or circles. Moreover, in some cases (e.g., no echo, $250$ MHz and $-200$ MHz) there are no crosses because the infidelity level of 0.3\% is never reached. However, for some detunings (e.g., $100$ MHz and $-130$ MHz) the duration is almost the same for the three considered levels of infidelity; as mentioned above, this indicates low leakage for relatively large drive amplitudes. There is even one weird case (no echo, $70$ MHz), where the order of the symbols is reversed (this is because of the unusual behavior of the dashed brown line in Fig.\ \ref{fig-echo} at the left, which relates to the increasing behavior of the orange line in Fig.\ \ref{fig-CNOT-time-num}).

As mentioned above, the difference between the symbols and the solid line in Fig.\ \ref{fig-cnot-delta-echo} is mainly determined by the leakage. We have checked that the main leakage channel for the detuning within the range $1/3 \alt \Delta/\eta_{\rm c}\alt 2/3$ is the transition $|0\rangle \to |2\rangle$ in the control qubit. This leakage channel is very strong when $\Delta/\eta_{\rm c}$ is close to $0.5$ (because of the resonance in the rotating frame), thus making impossible the practical operation of the CR gate at $\Delta/\eta_{\rm c}\simeq 0.5$. The leakage $|0\rangle \to |2\rangle$  becomes weaker for detunings farther away from this resonance point. For $\Delta/\eta_{\rm c}\agt 2/3$, the leakage channel $|1\rangle \to |2\rangle$ in the control qubit becomes more important (exact resonance at $\Delta/\eta_{\rm c}= 1$). Similarly, for $\Delta/\eta_{\rm c}\alt 1/3$, the leakage channel $|0\rangle \leftrightarrow |1\rangle$ in the control qubit becomes more important (exact resonance at $\Delta/\eta_{\rm c}= 0$). Thus, the leakage is relatively low for the detuning $\Delta$ near $(1/3)\eta_{\rm c}$ or $(2/3)\eta_{\rm c}$.
The trade-off between the lower leakage and shorter semi-analytical durations determines the best detuning ranges in Fig.\ \ref{fig-cnot-delta-echo}.

Note that besides the leakage channels $|0\rangle \to |2\rangle$, $|1\rangle \to |2\rangle$, and $|0\rangle \leftrightarrow |1\rangle$ for the control qubit, there are also important leakage channels, which involve level $|2\rangle$ of the target qubit. For example, for detunings of 210 and 230 MHz, the leakage is dominated by a near-resonance between levels $|01\rangle$ and $|12\rangle$. Also, for detunings of $-200$, $-170$, and $-130$ MHz, the main leakage is due to a near-resonance between levels $|11\rangle$ and $|02\rangle$. Overall, the interplay between different leakage channels is rather complicated, leading to a rather complicated behavior of numerical results in Fig.\ \ref{fig-cnot-delta-echo}.

Since the echo-CR gate is more affected by the leakage than the basic CR gate, a natural hypothesis is that its operation can be improved by using smoother ramps, in particular, by changing the relative duration of the front (and rear) ramps from the value $\tau_{\rm r}/\tau_{\rm p}=0.3$ (used in Fig.\ \ref{fig-cnot-delta-echo}) to the maximum possible value $\tau_{\rm r}/\tau_{\rm p}=0.5$. To check this hypothesis, we have also simulated the echo-CR gate operation with $\tau_{\rm r}/\tau_{\rm p}=0.5$, but the results were inconclusive: sometimes this makes CNOT time slightly shorter, sometimes slightly longer. For example, for the detuning of 190 MHz (our shortest-duration point), changing  $\tau_{\rm r}/\tau_{\rm p}$ from 0.3 to 0.5 increases the CNOT time by 4 ns for 0.3\% infidelity and by 1 ns for 1\% infidelity, but decreases it by 12 ns for 3\% infidelity. So, crudely, we think that the pulse shape ratio  $\tau_{\rm r}/\tau_{\rm p}=0.3$ is still reasonable for the echo-CR gate.

\bibliographystyle{apsrev}

\end{document}